\documentclass[preprintnumbers,article,amsmath,amssymb,floatfix,10pt,prd,superscriptaddress,nofootinbib]{revtex4-2}
\usepackage{bm}
\usepackage{amsfonts}
\usepackage{latexsym}
\usepackage[latin1]{inputenc}
\usepackage{graphicx}
\usepackage{amsmath}
\usepackage{palatino}
\usepackage{mathpazo}
\usepackage{textcomp}
\usepackage{float}
\usepackage{booktabs}
\usepackage{dcolumn}
\usepackage{ragged2e}
\usepackage{hyperref}
\hypersetup{colorlinks,citecolor=blue}
\hypersetup{colorlinks=true,linkcolor=red,filecolor=magenta,    urlcolor=blue}
\usepackage{amsmath}
\usepackage{xcolor}
\usepackage{orcidlink}
\usepackage{epsfig}
\usepackage{caption}
\usepackage{subcaption}
\usepackage{commath}
\def\jnl@style{\it}
\def\aaref@jnl#1{{\jnl@style#1}}

\def\aaref@jnl#1{{\jnl@style#1}}

\def\aj{\aaref@jnl{AJ}}                   
\def\apj{\aaref@jnl{ApJ}}                 
\def\apjl{\aaref@jnl{ApJ}}                
\def\apjs{\aaref@jnl{ApJS}}               
\def\apss{\aaref@jnl{Ap\&SS}}             
\def\aap{\aaref@jnl{A\&A}}                
\def\aapr{\aaref@jnl{A\&A~Rev.}}          
\def\aaps{\aaref@jnl{A\&AS}}              
\def\mnras{\aaref@jnl{Mon.~Not.~Roy.~Astron.~Soc.}}             
\def\prd{\aaref@jnl{Phys.~Rev.~D}}        
\def\prc{\aaref@jnl{Phys.~Rev.~C}}  
\def\prl{\aaref@jnl{Phys.~Rev.~Lett.}}    
\def\qjras{\aaref@jnl{QJRAS}}             
\def\skytel{\aaref@jnl{S\&T}}             
\def\ssr{\aaref@jnl{Space~Sci.~Rev.}}     
\def\zap{\aaref@jnl{ZAp}}                 
\def\nat{\aaref@jnl{Nature}}              
\def\aplett{\aaref@jnl{Astrophys.~Lett.}} 
\def\apspr{\aaref@jnl{Astrophys.~Space~Phys.~Res.}} 
\def\physrep{\aaref@jnl{Phys.~Rep.}}      
\def\physscr{\aaref@jnl{Phys.~Scr}}       
\def\commat{\aaref@jnl{Comm.~Math.~Phys.}}              
\def\science{\aaref@jnl{Science}}               
\def\cqg{\aaref@jnl{Classical Quant.~Grav.}}            
\def\jpcs{\aaref@jnl{JPCS}}                                     
\def\ijmpd{\aaref@jnl{Int.~J.~Mod.~Phys.~D}}                    
\def\grg{\aaref@jnl{Gen.~Relat.~Gravit.}}               
\def\rpp{\aaref@jnl{Rep.~Prog.~Phys.}}          
\def\npa{\aaref@jnl{Nucl.~Phys.~A}}        
\def\lrr{\aaref@jnl{Living Rev.~Rel.}}                   
\def\jcap{\aaref@jnl{J.~Cosmology Astropart.~Phys.}}    
\def\rmp{\aaref@jnl{Rev.~Mod.~Phys.}}   
\def\epjc{\aaref@jnl{Eur.~Phys.~J.~C}} 
\def\plb{\aaref@jnl{~Phy.~Lett.~B}} 
\def\mpla{\aaref@jnl{Mod.~Phy.~Lett.~A}} 
\def\arxiv{\aaref@jnl{arxiv.org}}

\allowdisplaybreaks[1]

\begin{document}
\color{black}       
\title{Observational constraints on two cosmological models of $f(Q)$ theory}

\author{M. Koussour\orcidlink{0000-0002-4188-0572}}
\email{pr.mouhssine@gmail.com}
\affiliation{Quantum Physics and Magnetism Team, LPMC, Faculty of Science Ben
M'sik,\\
Casablanca Hassan II University,
Morocco.}
\author{Avik De\orcidlink{0000-0001-6475-3085}}
\email{avikde@utar.edu.my}
\affiliation{Department of Mathematical and Actuarial Sciences, Universiti Tunku Abdul Rahman, Jalan Sungai Long, 43000 Cheras, Malaysia.}

%
\footnotetext{The research was supported by the Ministry of Higher Education (MoHE), through the Fundamental Research Grant Scheme (FRGS/1/2021/STG06/UTAR/02/1). }
\begin{abstract}
In the past few years, $f(Q)$ theories have drawn a lot of research attention in replacing Einstein's theory of gravity successfully. The current study
examines the novel cosmological possibilities emerging from two specific
classes of $f(Q)$ models using the parametrization form of the equation of
state (EoS) parameter as $\omega \left( z\right) =-\frac{1}{1+3\beta \left(
1+z\right) ^{3}}$, which displays quintessence behavior with the evolution
of the Universe. We do statistical analyses using the Markov chain Monte
Carlo (MCMC) method and background datasets like Type Ia Supernovae (SNe Ia)
luminosities and direct Hubble datasets (from cosmic clocks), and Baryon
Acoustic Oscillations (BAO) datasets. This lets us compare these new ideas about
the Universe to the $\Lambda $CDM model in a number of different possible
ways. We have come to the conclusion that, at the current level of accuracy,
the values of their specific parameters are the best fits for our $f(Q)$
models. To conclude the accelerating behavior of the Universe, we further study
the evolution of energy density, pressure, and deceleration parameter for
these $f(Q)$ models. 
\end{abstract}
\date{\today}
\maketitle

\section{Introduction}

\label{section 1} Despite the undeniable success of the `standard model of
cosmology', governed by general relativity (GR), there are atleast two
motivations behind the popularity of the modified theories of gravity in
modern research. The first being to alleviate the requirement of dark sector
to reason the present acceleration of the Universe. The other is purely
theoretical, towards the renormalizability of GR and to extend it to an
ultimate theories of quantum gravity. Teleparallel gravity has been
extensively researched in this regard in the past few years. To be specific,
the Levi-Civita connection, which serves as the foundation of GR, may be
replaced by a general affine connection on spatially flat spacetime with
vanishing torsion, allowing its non-metricity to assume complete
responsibility for defining gravity. `Symmetric teleparallel gravity' is the
name given to this particular theory. Long back, in order to unify gravity
with electromagnetism, Einstein \cite{1} attributed gravity to the torsion
of spacetime in a more developed `metric teleparallel theory', which is
based on an affine connection with vanishing curvature and non-metricity
both. Under teleparallel theories, one can construct either the so-called
torsion scalar $\mathbb{T}$ from this torsion tensor in the metric
teleparallel branch or the non-metricity scalar $Q$ in its symmetric
counterpart. Thereafter, in the Einstein-Hilbert action term of GR, one can
replace the Ricci scalar by $\mathbb{T}$ in the metric teleparallel theory
and by $Q$ in the symmetric teleparallel theory to obtain the respective
field equations. However, other than a total divergence term, the theories
are actually equivalent to GR and thus also rely on dark components of the
Universe as GR does. To address this dark sector issue of the Universe,
without inviting a scalar field, natural extensions in terms of $f(\mathbb{T}%
)$ and $f(Q)$-theories were motivated. While these three theories of gravity (often referred to as the geometrical trinity of gravity) are regarded to be entirely equivalent, these two counterparts of GR nearly went ignored until recently, when the extension in these two theories, respectively $f(\mathbb{T})$ and $f(Q)$ theories of gravity in line with the $f(R)$ extension of GR were studied as a possible alternate source of dark energy to beautifully explain the present accelerating Universe \cite{accfT1,accfT2,ko1,ko2,ko3,ko4,ko5,accfQ1,accfQ2,accfQ3}. The foremost advantage of these two theories are their second
order field equations, as in GR, and unlike in the modified $f(R)$ gravity theory where the field equation is of the fourth order \cite{63}. However, the older and much matured $f(\mathbb{T})$ formulation displays certain issues as addressed in an array of publications \cite{fT/issue, fT/issue1, fT/issue2, fT/lli} which the $f(Q)$ theory is free from. 
On generic FLRW backgrounds, the significant coupling difficulties that may be observed in $f(\mathbb{T})$ theories are absent in $f(Q)$ models. Also, the predictions of the $f(Q)$ and $f(\mathbb{T})$ models correspond in the small-scale quasi-static limit, but that at higher scales the $f(Q)$ models generically transmit 2 scalar degrees of freedom that are absent in the case of $f(\mathbb{T})$. These two degrees of freedom vanish around maximally symmetric backgrounds, resulting
in the strong coupling problem that has been explored \cite{61}. In addition, one can investigate the fascinating work \cite{62}, in which the formalism of $f(\mathbb{T})$ gravity was created in agreement with GR. In a further serious scrutiny of the field equations of $f(\mathbb{T})$ theories we can observe an unusual presence of anti-symmetric elements, which is not the case in $f(Q)$ or $f(R)$ or most other theories of gravity.  This anti-symmetric part of the field equation of $f(\mathbb{T})$ obtained from the variation of the action term with respect to the metric tensor actually equals the field equation obtained from the variation of the same action by the affine connection \cite{ad/bianchi}. As a result, $f(\mathbb{T})$ theories (except $f(\mathbb{T})=\mathbb{T}$) are also not locally Lorentz invariant and possess extra degrees of freedom which remain absent from GR. This is a devastating blow, as the absence of Lorentz invariance necessitates the implementation of a system of $16$ equations in $f(\mathbb{T})$ theories instead of $10$ equations in GR. On a side by side comparison, the $f(\mathbb{T})$ connections must comply with $Q_{\mu\nu\gamma}=0$, which are $40$ independent equations due to the symmetry of the non-metricity tensor $Q_{\mu\nu\gamma}$ in the second and third indices. Whereas, the $f(Q)$ connections must obey the vanishing torsion tensor criteria, $\mathbb{T}^\gamma_{\,\,\,\mu\nu}=0$; only $24$ independent equations since the $\mu=\nu$ case is trivially satisfied, making $f(\mathbb{T})$ connections much more restrictive than $f(Q)$. Consequently, in a spatially flat FLRW spacetime, there are in total 3 classes of $f(Q)$ formulations possible, yielding 3 different aspects of cosmology under the umbrella of the same $f(Q)$ theory \cite{flrwfQ}. Whereas, there is only a single class of $f(\mathbb{T})$ formulation in that spacetime, and this class is predominantly equivalent to the one obtained using the diagonal tetrad in the Cartesian coordinates \cite{flrwfT}. For an in depth comparison of these two theories, one can further look into \cite{fQfT, fQfT1, fQfT3, gde, fQfT2, de/epjc}. 
However, as mentioned earlier, the modified $f(Q)$ theory under the symmetric teleparallelism is merely at its infancy; a lot of theoretical investigation is still due before we can vouch for its robustness. Specially, the newly discovered $f(Q)$ constructions arising from the non-vanishing classes of affine connections in both isotropic and homogeneous spacetimes \cite{flrwfQ} and static spherically symmetric spacetimes \cite{bhfQ} are yet to be tested.


On the other hand, the ability to formulate practical guidelines for cosmological applications gives birth to the relationship between theory and observational datasets. For a certain class of $f(Q)$ gravity theories under symmetric teleparallelism, the Hubble parameter may be computed analytically by simply adhering to some partial extensions of generally accepted conventional approaches. 
In this regard, we should mention some of the introductory data analysis works conducted in $f(Q)$ theory. In \cite{lcdm}, $f(Q)$ theory challenged the $\Lambda$CDM model for the first time, in the sense that the new gravity theory despite having the same number of free parameters as in $\Lambda$CDM, at a
cosmological framework it possibly can avoid $\Lambda$CDM as a limit and thus can alleviate the cosmological constant problem. Moreover, contact with observations at both background and perturbation levels reveals that the model, in some datasets is slightly preferred than $\Lambda$CDM cosmology, although
in all cases the two models are statistically indiscriminate. The $f(Q)$ model also does not exhibit early dark energy features, and thus it immediately passes BBN constraints. Within few weeks of this work, support in the form of \cite{lcdm1} appeared.
Modern cosmology probes are used in the statistical analysis, and a certain matter-energy composition is assumed, such that the candidate models provide promising modified gravity candidates to represent the cosmic backdrop. Starting with an expression for the pressure-energy density ratio, the so-called equation of state (EoS) parameter $\omega$, the technique moves on to the derivation and analysis of other cosmic parameters corresponding to two particular $f(Q)$ models, namely, $f(Q)=-Q+%
\frac{\alpha}Q$ and $f(Q)= -\alpha Q^n$, both of which demonstrate excellent
fit with the cosmological data. The model parameters $\alpha$ and $n$ can be
easily adjusted to retrieve GR, providing the opportunity for a clean comparison.

The present paper is organised as follows:\\
After the brief introductory discussion presented above, in Section \ref{sec2} the fundamental mathematical formulation of the modified $f(Q)$ theories is presented, followed by the (effective) pressure and energy equations in the spatially flat Friedmann-Lema\^itre-Robertson-Walker (FLRW) geometric background in Section \ref{sec3}. We consider a form of the EoS parameter $\omega(z)$ as a function of red-shift parameter $z$ in Section \ref{section 4}. Under this section, two separate $f(Q)$ models are analysed in two separate subsections and the required cosmological parameters are derived. Subsection \ref{subsecA} assumes $f(Q)=-Q+\frac{\alpha}{Q}$ and Subsection \ref{subsecB} assumes the model $f(Q)=-\alpha Q^n$. Section \ref{section 5} delivers all the relevant observational data analysis of the above-mentioned models. A detailed concluding remarks on all the acquired findings in Section \ref{section 7} is preceded by Section \ref{section 6} discussing briefly the behavior of the cosmological parameters.  


\section{Basic concepts of $f(Q)$ gravity}\label{sec2}

In this section We discuss the detailed formulation of the symmetric
teleparallelism, specially its extension the modified $f(Q)$ theory. We
begin with a $4$-dimensional Lorentzian manifold $M^{4}$, a line element
governed by the metric tensor $g_{\mu \nu }$ in certain coordinate system $%
\{x^{0},x^{1},x^{2},x^{3}\}$ and a non-tensorial term, the affine connection 
$\Gamma _{\,\,\,\mu \nu }^{\alpha }$, defining the covariant derivative $%
\nabla $ and also taking care of the three main aspects of the spacetime
geometry corresponding to this connection, the curvature, torsion and
non-metricity. However, once we restrict ourselves to specifically consider
vanishing of both the non-metricity and the torsion tensors corresponsing to
the connection, we can assert that there is only a unique connection
available, the Levi-Civita connection $\mathring{\Gamma}$ and it has a
well-known relation with the metric $g$ given by 
\begin{equation}
\mathring{\Gamma}_{\,\,\,\mu \nu }^{\alpha }=\frac{1}{2}g^{\alpha \beta
}\left( \partial _{\nu }g_{\beta \mu }+\partial _{\mu }g_{\beta \nu
}-\partial _{\beta }g_{\mu \nu }\right) .
\end{equation}

So, the Levi-Civita connection is basically a function of the metric $g$ and
not an independent player in the spacetime geometry. Here, instead we
consider a torsion-free affine connection $\Gamma $ on a flat spacetime
which is not metric-compatible, the incompatibility is characterised by the
non-metricity tensor 
\begin{equation}
Q_{\lambda \mu \nu }=\nabla _{\lambda }g_{\mu \nu }=\partial _{\lambda
}g_{\mu \nu }-\Gamma _{\,\,\,\lambda \mu }^{\beta }g_{\beta \nu }-\Gamma
_{\,\,\,\lambda \nu }^{\beta }g_{\beta \mu }\neq 0\,,  \label{Q tensor}
\end{equation}

We can always express 
\begin{equation}
\Gamma ^{\lambda }{}_{\mu \nu }=\mathring{\Gamma}^{\lambda }{}_{\mu \nu
}+L^{\lambda }{}_{\mu \nu }  \label{connc}
\end{equation}%
where $L^{\lambda }{}_{\mu \nu }$ is the disformation tensor. It follows
that 
\begin{equation}
L^{\lambda }{}_{\mu \nu }=\frac{1}{2}(Q^{\lambda }{}_{\mu \nu }-Q_{\mu
}{}^{\lambda }{}_{\nu }-Q_{\nu }{}^{\lambda }{}_{\mu })\,.  \label{L}
\end{equation}

We can construct two different types of non-metricity vectors, 
\begin{equation}
Q_{\mu }=g^{\nu \lambda }Q_{\mu \nu \lambda }=Q_{\mu }{}^{\nu }{}_{\nu
}\,,\qquad \tilde{Q}_{\mu }=g^{\nu \lambda }Q_{\nu \mu \lambda }=Q_{\nu \mu
}{}^{\nu }\,.
\end{equation}

The non-metricity conjugate tensor $P^{\lambda }{}_{\mu \nu }$ is given by 
\begin{equation}
P^{\lambda }{}_{\mu \nu }=\frac{1}{4}\left( -2L^{\lambda }{}_{\mu \nu
}+Q^{\lambda }g_{\mu \nu }-\tilde{Q}^{\lambda }g_{\mu \nu }-\frac{1}{2}%
\delta _{\mu }^{\lambda }Q_{\nu }-\frac{1}{2}\delta _{\nu }^{\lambda }Q_{\mu
}\right) \,.  \label{P}
\end{equation}

Finally, one can define the non-metricity scalar 
\begin{equation}
Q=-Q_{\lambda \mu \nu }P^{\lambda \mu \nu }=\frac{1}{2}Q_{\lambda \mu \nu
}L^{\lambda \mu \nu }+\frac{1}{2}Q_{\lambda }\tilde{Q}^{\lambda }-\frac{1}{4}%
Q_{\lambda }Q^{\lambda }.  \label{Q}
\end{equation}

By varying the action term 
\begin{equation}
S=\int \left[ \frac{1}{2\kappa }f(Q)+\mathcal{L}_{M}\right] \sqrt{-g}\,d^{4}x
\label{action}
\end{equation}%
with respect to the metric tensor $g^{\mu \nu }$, we obtain the field
equation \cite{coincident}
\begin{equation}
\frac{2}{\sqrt{-g}}\nabla _{\lambda }(\sqrt{-g}f_{Q}P^{\lambda }{}_{\mu \nu
})+\frac{1}{2}fg_{\mu \nu }+f_{Q}(P_{\nu \rho \sigma }Q_{\mu }{}^{\rho
\sigma }-2P_{\rho \sigma \mu }Q^{\rho \sigma }{}_{\nu })=-\kappa T_{\mu \nu
}.  \label{FE1}
\end{equation}%
$T_{\mu \nu }$ is the energy-momentum tensor generated from the matter
Lagrangian $\mathcal{L}_{M}$. We assume a barotropic perfect fluid given by 
\begin{equation*}
T_{\mu \nu }=(p+\rho )u_{\mu }u_{\nu }+pg_{\mu \nu },
\end{equation*}%
with isotropic pressure $p$, energy-density $\rho $ and the four-velocity
vector $u^{\mu }$. Recently, the covariant representation of the field
equation (\ref{FE1}) was derived \cite{zhao} 
\begin{equation}
f_{Q}\mathring{G}_{\mu \nu }+\frac{1}{2}g_{\mu \nu }(f-Qf_{Q})+2f_{QQ}%
\mathring{\nabla}_{\lambda }QP^{\lambda }{}_{\mu \nu }=-\kappa T_{\mu \nu },
\label{FE2}
\end{equation}%
where $\mathring{G}_{\mu \nu }=\mathring{R}_{\mu \nu }-\frac{1}{2}g_{\mu \nu
}\mathring{R}$, is the Einstein tensor corresponding to the Levi-Civita
connection. The field equations (\ref{FE2}) can be equivalently written in
the effective form 
\begin{equation}
\mathring{G}_{\mu \nu }=-\frac{\kappa }{f_{Q}}T_{\mu \nu }^{eff}=-\frac{%
\kappa }{f_{Q}}T_{\mu \nu }+T_{\mu \nu }^{DE},  \label{effenergy}
\end{equation}%
where $\frac{-\kappa }{f_{Q}}$ is the effective gravitational constant and
for its positivity in our construction, we assume $f_{Q}<0$. The dark energy
component emerged from this modification of STEGR is given by 
\begin{equation}
T_{\mu \nu }^{DE}=\frac{-1}{f_{Q}}[\frac{1}{2}g_{\mu \nu }(f-Qf_{Q})+2f_{QQ}%
\mathring{\nabla}_{\lambda }QP^{\lambda }{}_{\mu \nu }].  \label{de}
\end{equation}

On the other hand, since the affine connection is an independent entity in
the symmetric teleparallel theory, we vary the action (\ref{action}) with
regard to the affine connection $\Gamma $ to obtain the connection field
equation as \cite{zhao} 
\begin{equation}
\nabla _{\mu }\nabla _{\nu }(\sqrt{-g}f_{Q}P^{\mu \nu }{}_{\gamma })=0,
\label{2q}
\end{equation}%
on the basis of the assumption that the matter Lagrangian $\mathcal{L}_{M}$
is not a function of the affine connection. Moreover, it has been showed
that \cite{ad/bianchi} this second field equation (\ref{2q}) is trivially
satisfied in a model-independent manner in the spacetime geometry we are
going to consider in the Section \ref{sec3}. So in the present article
our sole attention is devoted to the metric field equation (\ref{FE2}). It
should be mentioned that $R_{\sigma \mu \nu }^{\rho }=0$ is one of the
restrictions that was utilised when developing the $f(Q)$-theory. This
indicates that there is a unique coordinate system that can be chosen to
make the affine connection disappears, denoted by the expression $\Gamma
_{\mu \nu }^{\lambda }=0$. The term \textquotedblleft coincident gauge"
refers to this kind of circumstance. For more details of the $f(Q)$ theory and its cosmological applications, one can see \cite{perturb, Dimakis, Beh, lcdm, Q3, ad/ec, lin, cosmography, signa, red-shift, dynamical1} and the references therein. 


\section{$f(Q)$ Cosmology in isotropic and homogeneous Universe}\label{sec3}

The "cosmological principle" states that on a large enough scale our
Universe is homogenous and isotropic, that is, it is the same at every point
and in every direction. Based on this, the most reasonable and theoretically
and observationally supported model of the present Universe is the spatially
flat Friedmann-Lemaitree-Robertson-Walker (FLRW) spacetime given by the line
element in Cartesian coordinates 
\begin{equation}
ds^{2}=-dt^{2}+a^{2}(t)[dx^{2}+dy^{2}+dz^{2}],  \label{3a}
\end{equation}%
where $a(t)$ is the scale factor of the Universe. We proceed with the
coincident gauge choice as discussed above and obtain the non-metricity
scalar as $Q=6H^{2}$, where $H=\frac{\overset{.}{a}}{a}$ is the Hubble
parameter, which measures the expansion rate of the Universe, and $\dot{()}$
indicates a derivative with regard to cosmic time $t$.

In this context, the field equations (\ref{FE2}) give the following
expressions of the energy density $\rho $ and the isotropic pressure $p$ 
\cite{zhao,ad/ec} 
\begin{equation}
\rho =\frac{f}{2}-6H^{2}f_{Q},  \label{F22}
\end{equation}%
\begin{equation}
p=\left( \overset{.}{H}+\frac{\overset{.}{f_{Q}}}{f_{Q}}H\right) \left(
2f_{Q}\right) -(\frac{f}{2}-6H^{2}f_{Q}),  \label{F33}
\end{equation}

Furthermore, to explain cosmic history and the possible transition to an
accelerated phase, we use the equation of state (EoS) parameter $\omega $,
given by 
\begin{equation}
\omega =\frac{p}{\rho }=-1+\frac{\left( \overset{.}{H}+\frac{\overset{.}{%
f_{Q}}}{f_{Q}}H\right) \left( 2f_{Q}\right) }{\left( \frac{f}{2}%
-6H^{2}f_{Q}\right) }.  \label{EoS}
\end{equation}

On the other hand, using (\ref{effenergy}) the effective energy density $%
\rho _{eff}$ and effective pressure $p_{eff}$ of the cosmic fluid can be
written as \cite{ad/ec} 
\begin{align}
3H^{2}&=-\frac{1}{f_{Q}}\rho _{eff}\,=-\frac{1}{f_{Q}}\left( \rho +\frac{%
6H^{2}f_{Q}-f}{2}\right) ,  \label{effrho} \\
-(2\dot{H}+3H^{2})&=-\frac{1}{f_{Q}}p_{eff}\,=-\frac{1}{f_{Q}}\left( p+\frac{%
f-6H^{2}f_{Q}-4\overset{.}{f}_{Q}H}{2}\right) .  \label{effp}
\end{align}

Further, in the limiting situation $f\left( Q\right) =-Q=-6H^{2}$, the
gravitational action (\ref{action}) is reduced to the standard
Hilbert-Einstein form and Equations (\ref{effrho}) and (\ref{effp}) reduce
to the standard Friedmann equations of GR, $3H^{2}=\rho $, and $2\dot{H}%
+3H^{2}=-p$, respectively. Thus, the effective EoS parameter $\omega _{eff}$%
\ is%
\begin{equation}
\omega _{eff}=\frac{p_{eff}}{\rho _{eff}}=\frac{2p+f-6H^{2}f_{Q}-4\overset{.}%
{f}_{Q}H}{2\rho +6H^{2}f_{Q}-f}.
\end{equation}

To describe the accelerated/decelerated aspect of the cosmic expansion, we
consider the deceleration parameter $q$, expressed as,%
\begin{equation}
q=-\frac{\overset{.}{H}}{H^{2}}-1,  \label{q}
\end{equation}


\section{Cosmological models with specific form of EoS}\label{section 4}

In this section, we examine dark energy parametrization, which displays
quintessence behavior with the evolution of the cosmos. Our principal
objective is to study this parametrization using existing cosmological data.
For simplicity, we will adopt the red-shift as the independent variable,
expressed as $z=\frac{a_{0}}{a}-1$, with the current scale factor $a_{0}$
fixed to $1$. In general, there is no theoretical method for selecting the
optimal $\omega \left( z\right) $, but by utilizing observational data,
suitable parametrizations can be found. In literature, several EoS dark
energy parametrization models were proposed and fitted with observational
data. Ref. \cite{P1} proposed an one-parameter family of EoS dark energy
model. Two-parameters family of EoS dark energy parametrizations, especially
the Chevallier-Polarski-Linder parametrization \cite{P2, P3}, the Linear
parametrization \cite{P3, P4, P5, P6}, the Logarithmic parametrization \cite%
{P7}, the Jassal-Bagla-Padmanabhan parametrization \cite{P8}, and the
Barboza-Alcaniz parametrization \cite{P9}, were also investigated. Further,
in \cite{P10, P11, P12} three and four parameters family of EoS dark energy
parametrizations are examined. Here, we assume that the EoS parameter is
parametrized as a function of red-shift $z$,%
\begin{equation}
\omega \left( z\right) =-\frac{1}{1+3\beta \left( 1+z\right) ^{3}},
\label{eff}
\end{equation}%
where $\beta $ is the only free parameter. The purpose of choosing this
parametrization for $\omega \left( z\right) $ is that for very large
red-shift $z\gg 1$, i.e., at the early phases of cosmological evolution, $%
\omega $ is approximately zero, representing the behavior of the EoS
parameter for a pressureless fluid (ordinary matter), but it gradually
decreases to negative values at the present i.e. $z=0$, leads to negative
pressure and $\omega =-\frac{1}{1+3\beta }$. This last equation clearly
shows that for an accelerated Universe scenario: $\omega \leq -\frac{1}{3}$,
resulting in a constraint of the value of $\beta $ as $\beta \leq \frac{2}{3}
$. In addition, we can see that at later times $z\rightarrow -1$, $\omega $
tends to $-1$, which is similar to the behavior of the cosmological constant 
$\Lambda $. Also, for $\beta =0$, $\omega $ reduces to $-1$. These
situations are summarized below,

\begin{itemize}
\item $\omega \longrightarrow -1$, as $z\longrightarrow -1$,

\item $\omega \longrightarrow 0$, for $z\gg 1$,

\item $\omega =-\frac{1}{1+3\beta }$, for $z=0$.
\end{itemize}

To make comparisons of theoretical results with cosmological data simple, we
utilize the red-shift $z$ instead of the usual time variable $t$. As a
result, we can change the derivatives with regard to time with the
derivatives with regard to red-shift using the relationship,%
\begin{equation}
\frac{d}{dt}=\frac{dz}{dt}\frac{d}{dz}=-(1+z)H(z)\frac{d}{dz}.
\end{equation}

The deceleration parameter $q$ can be calculated as a function of cosmic
red-shift,%
\begin{equation}
q\left( z\right) =\left( 1+z\right) \frac{1}{H\left( z\right) }\frac{%
dH\left( z\right) }{dz}-1.  \label{qz}
\end{equation}

Also, the derivative of the Hubble parameter can be expressed as, 
\begin{equation}
\dot{H}=-(1+z)H(z)\frac{dH}{dz}.
\end{equation}

Now, we examine several specific cosmological models in $f(Q)$ gravity
theory, models that correspond to different classes of the function $f(Q)$.
We also analyze the behavior of geometric and physical cosmological
parameters in $f(Q)$ gravity such as energy density, pressure, and
deceleration parameter.

\subsection{$f\left( Q\right) =-Q+\frac{\protect\alpha }{Q}$}\label{subsecA}

For a first case of a cosmological model in $f(Q)$ gravity, consider the
scenario where the function $f(Q)$ can be expressed as, $f(Q)=-Q+\frac{%
\alpha }{Q}$, where $\alpha $ is a constant. So, we get $f_{Q}=-1-\frac{%
\alpha }{Q^{2}}$ and $f_{QQ}=\frac{2\alpha }{Q^{3}}$. The Friedmann
equations \eqref{F22} and \eqref{F33} for this particular $f(Q)$ model,
reduce to%
\begin{equation}
\rho =\frac{\alpha }{4H^{2}}+3H^{2},
\end{equation}%
and%
\begin{equation}
p=\frac{\alpha \overset{.}{H}}{6H^{4}}-2\overset{.}{H}-\frac{\alpha }{4H^{2}}%
-3H^{2}.
\end{equation}

Using (\ref{EoS}), we obtain the EoS parameter in terms of Hubble parameter
as 
\begin{equation}
\omega =\frac{2\overset{.}{H}}{3H^{2}}-\frac{16\overset{.}{H}H^{2}}{\alpha
+12H^{4}}-1.  \label{17}
\end{equation}

The differential equation for $H\left( z\right) $ is obtained by (\ref{17})
the presumed ansatz of $\omega $\ as indicated in (\ref{eff}),%
\begin{equation}
\frac{2\overset{.}{H}\left( \alpha -12H^{4}\right) }{3H^{2}\left( \alpha
+12H^{4}\right) }-\frac{3\beta (z+1)^{3}}{3\beta (z+1)^{3}+1}=0.
\end{equation}

Therefore, the solution obtained for the Hubble parameter $H\left( z\right) $
as a function of red-shift $z$\ is 
\begin{equation}
H\left( z\right) =\frac{\sqrt{\sqrt{\frac{\left( \alpha +12H_{0}^{4}\right)
^{2}\left( 3\beta (z+1)^{3}+1\right) ^{2}}{(3\beta +1)^{2}H_{0}^{4}}%
-48\alpha }+\frac{\left( \alpha +12H_{0}^{4}\right) \left( 3\beta
(z+1)^{3}+1\right) }{(3\beta +1)H_{0}^{2}}}}{2\sqrt{6}},  \label{H1}
\end{equation}%
where $H_{0}$ represents the current value (i.e. at $z=0$) of the Hubble
parameter.

For this specific case, the expression of the deceleration parameter
obtained by including (\ref{H1}) into (\ref{qz}) as%
\begin{equation}
q\left( z\right) =\frac{9\beta (z+1)^{3}\left( \alpha +12H_{0}^{4}\right) }{%
2(3\beta +1)H_{0}^{2}\sqrt{\frac{\left( \alpha +12H_{0}^{4}\right)
^{2}\left( 3\beta (z+1)^{3}+1\right) ^{2}}{(3\beta +1)^{2}H_{0}^{4}}%
-48\alpha }}-1.  \label{qz1}
\end{equation}

\subsection{$f\left( Q\right) =-\protect\alpha Q^{n}$}\label{subsecB}

For a second case of a cosmological model in $f(Q)$ gravity, we assume the
scenario where the function $f(Q)$ can be expressed as a power-law form, $%
f(Q)=-\alpha Q^{n}$, where $\alpha $ and $n$ are model parameters. So, we
get $f_{Q}=-\alpha nQ^{n-1}$ and $f_{QQ}=-\left( n-1\right) \alpha nQ^{n-2}$%
. The Friedmann equations \eqref{F22} and \eqref{F33} for this particular $%
f(Q)$ model, reduce to%
\begin{equation}
\rho =\alpha 2^{n-1} 3^n (2 n-1) \left(H^2\right)^n,
\end{equation}%
and%
\begin{equation}
p=\alpha \left(-6^{n-1}\right) (2 n-1) \left(H^2\right)^{n-1} \left(2\overset%
{.}{H} n+3 H^2\right).
\end{equation}

Using (\ref{EoS}), we obtain the EoS parameter in terms of Hubble parameter
as 
\begin{equation}
\omega =-\frac{2\overset{.}{H}n}{3H^{2}}-1.  \label{17}
\end{equation}

The differential equation for $H\left( z\right) $ is obtained by Eq. (\ref%
{17}) the presumed ansatz of $\omega $\ as indicated in Eq. (\ref{eff}),%
\begin{equation}
\frac{2\overset{.}{H}n}{3H^{2}}+\frac{3\beta (1+z)^{3}}{3\beta (1+z)^{3}+1}=0.
\end{equation}

Therefore, the solution obtained for the Hubble parameter $H\left( z\right) $
as a function of red-shift $z$\ is, 
\begin{equation}
H\left( z\right) =H_{0}\left[ \frac{3\beta (1+z)^{3}+1}{3\beta +1}\right] ^{%
\frac{1}{2n}},  \label{H2}
\end{equation}%
where $H_{0}$ represents the current value (i.e. at $z=0$) of the Hubble
parameter. The previous equation can be rewritten as,%
\begin{equation}
H\left( z\right) =H_{0}\left[ \left( 1-\gamma \right) (1+z)^{3}+\gamma %
\right] ^{\frac{1}{2n}},
\end{equation}

For reasons of simplicity, we have introduced $\gamma =\frac{1}{1+3\beta }$.
It should be observed that the standard $\Lambda $CDM model is equivalent to
the scenario $n=1$, with the current cold dark matter density parameter $%
\Omega _{m}^{0}=(1-\gamma) $. As a result, the model parameter $n$ is an
excellent indicator of the current model's deviation from the $\Lambda $CDM
model.

For this specific case, the expression of the deceleration parameter
obtained by including Eq. (\ref{H2}) into Eq. (\ref{qz}) as,%
\begin{equation}
q\left( z\right) =\frac{9\beta (1+z)^{3}}{6\beta n(1+z)^{3}+2n}-1.
\label{qz1}
\end{equation}

In the next section, we will attempt to estimate the values of $H_{0}$, $n$, $\alpha$, and $\beta$ using Hubble, SNe Ia, and BAO datasets. The behavior of cosmological parameters such as density, pressure, and deceleration parameter can be examined for each model (Models 1 and 2) using the values of $H_{0}$, $n$, $\alpha$, and $\beta$.

\section{Observational constraints from Hubble, SNe Ia and BAO}\label{section 5}

This section is involved with the different observational datasets used to
restrict the parameters $\alpha $, $\beta $, $n$ and $H_{0}$. 
To get the posterior distributions of the parameters, we use the usual
Bayesian algorithms and a Markov Chain Monte Carlo (MCMC) approach with the
emcee python package \cite{Mackey/2013}. This stimulation is accomplished by
the use of Hubble measurements (i.e., Hubble datasets), Type Ia supernovae
(SNe Ia) datasets, and BAO datasets. The probability function is used to
maximize the best fits of the parameters%
\begin{equation}
\mathcal{L}\propto exp(-\chi ^{2}/2),
\end{equation}%
where $\chi ^{2}$ is the chi-square function. The $\chi ^{2}$ functions are
explained below for different datasets.

\subsection{Hubble datasets}

The Hubble results are the first observational data sample used during our
research. The Hubble parameter is written as $H(z)=-dz/[dt(1+z)]$. Given
that $dz$ is determined via a spectroscopic survey, the model-independent
value of the $H(z)$ can be estimated by measuring $dt$. It is commonly known
that the Hubble parameter can directly estimate the rate of cosmological
expansion. In principle, two methods for calculating the Hubble parameter at
various red-shifts are widely used: the differential ages $\Delta t$\ of
galaxies and the line of sight BAO technique. In this work, we restrict the
model using a collection of $57$ Hubble parameter data points in the
red-shift range $0.07\leq z\leq 2.41$ published by Sharov and Vasiliev \cite%
{Hubble}. For Hubble datasets, the $\chi _{Hubble}^{2}$ function is given
respectively for the two models 
\begin{equation}
\chi _{Hubble\_1}^{2}\left( H_{0},\alpha ,\beta \right)
=\sum\limits_{k=1}^{57}\frac{[H_{th}(z_{k},H_{0},\alpha ,\beta
)-H_{obs}(z_{k})]^{2}}{\sigma _{H(z_{k})}^{2}},
\end{equation}%
and%
\begin{equation}
\chi _{Hubble\_2}^{2}\left( H_{0},n,\beta \right) =\sum\limits_{k=1}^{57}%
\frac{[H_{th}(z_{k},H_{0},n,\beta )-H_{obs}(z_{k})]^{2}}{\sigma
_{H(z_{k})}^{2}}.
\end{equation}

Here, $H_{obs}$ is the Hubble parameter value recovered from cosmic
observations, $H_{th}$ is its theoretical value estimated, and $\sigma _{H}$
is the standard deviation in the observed value of $H\left( z\right) $. The $%
1-\sigma $ and $2-\sigma $ contour graphs in Figs. \ref{H1} and \ref{H2}
show the best fit values for the parameters of both models obtained from the
Hubble datasets. The likelihoods are extremely closely adapted to Gaussian
distributions. The best-fit values of the model parameters found are: $%
H_{0}=69.3_{-2.2}^{+2.4}$, $\alpha =3.9_{-4.0}^{+5.2}$, and $\beta
=0.129_{-0.024}^{+0.027}$ for the first model, and $H_{0}=65.5_{-4.6}^{+4.5}$%
, $n=1.16_{-0.15}^{+0.14}$, and $\beta =0.33_{-0.24}^{+0.39}$ for the second
model. Also, Figs. \ref{ErrorHubble1} and \ref{ErrorHubble2} show the error
bar plots for both models and the $\Lambda $CDM, with the cosmological
constant density parameter $\Omega _{\Lambda }^{0}=0.7$, the matter density
parameter $\Omega _{m}^{0}=0.3$ and $H_{0}=69$ km/s/Mpc. It is shown that
the $f(Q)$ models completely suit the observational data while deviating
somewhat from the $\Lambda $CDM.

\begin{figure}[!htb]
\begin{minipage}{0.49\textwidth}
     \centering
   \includegraphics[scale=0.6]{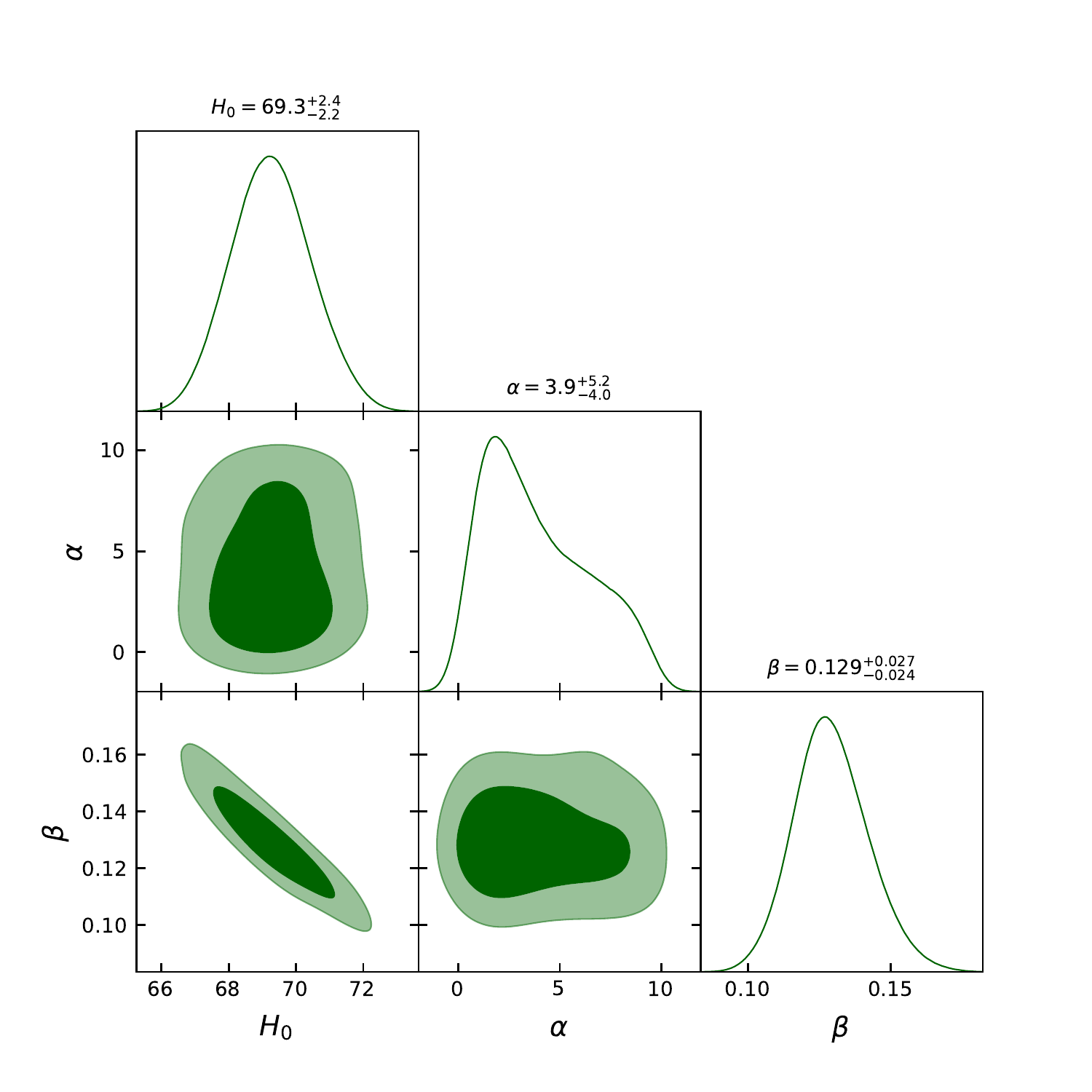}
\caption{Constraints on the model parameters at $1-\protect\sigma $ and $2-%
\protect\sigma $ confidence interval using the Hubble datasets (Model 1).}\label{H1}
   \end{minipage}\hfill 
\begin{minipage}{0.49\textwidth}
     \centering
    \includegraphics[scale=0.6]{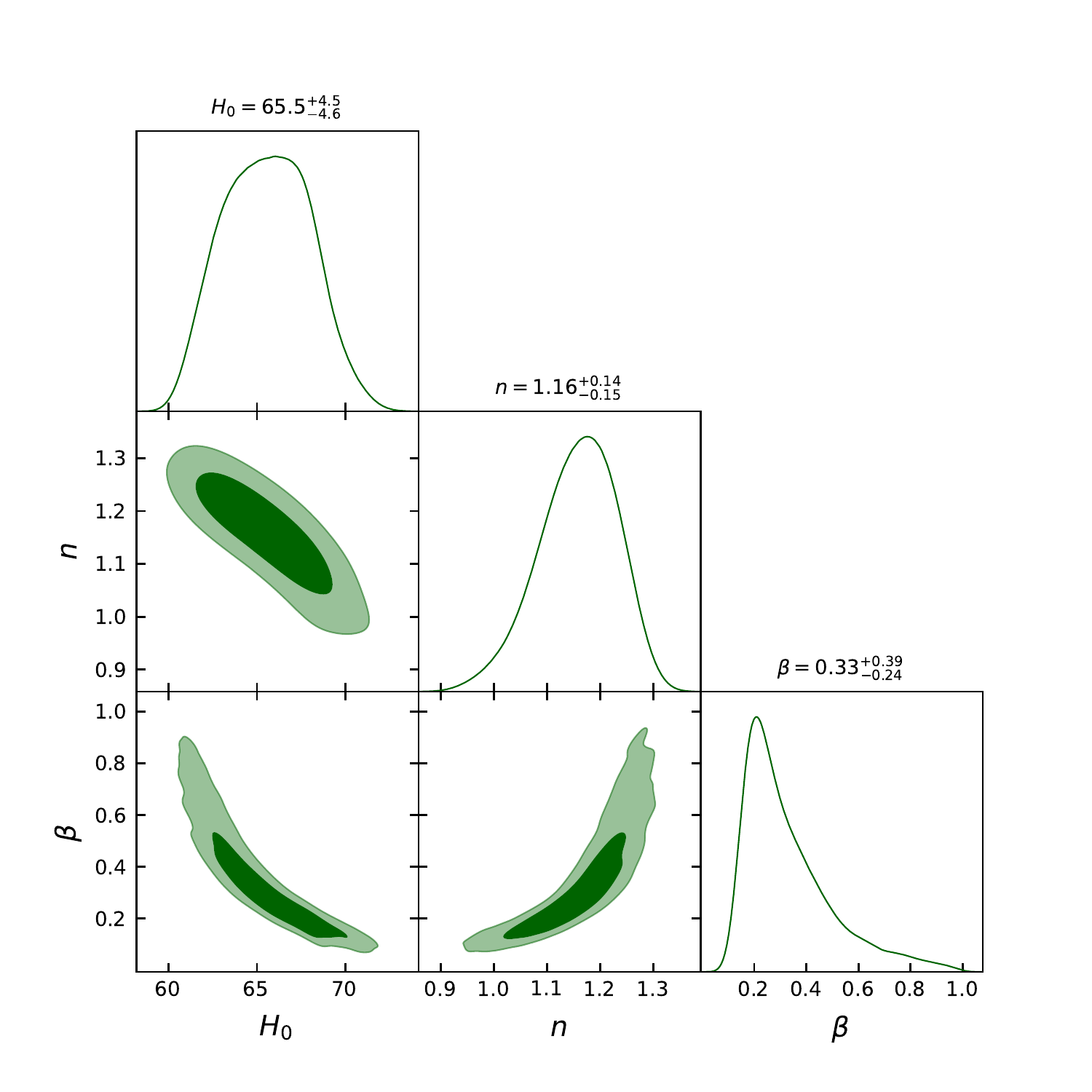}
\caption{Constraints on the model parameters at $1-\protect\sigma $ and $2-%
\protect\sigma $ confidence interval using the Hubble datasets (Model 2).}\label{H2}
   \end{minipage}
\end{figure}

\begin{widetext}

\begin{figure}[h]
\centerline{\includegraphics[scale=0.60]{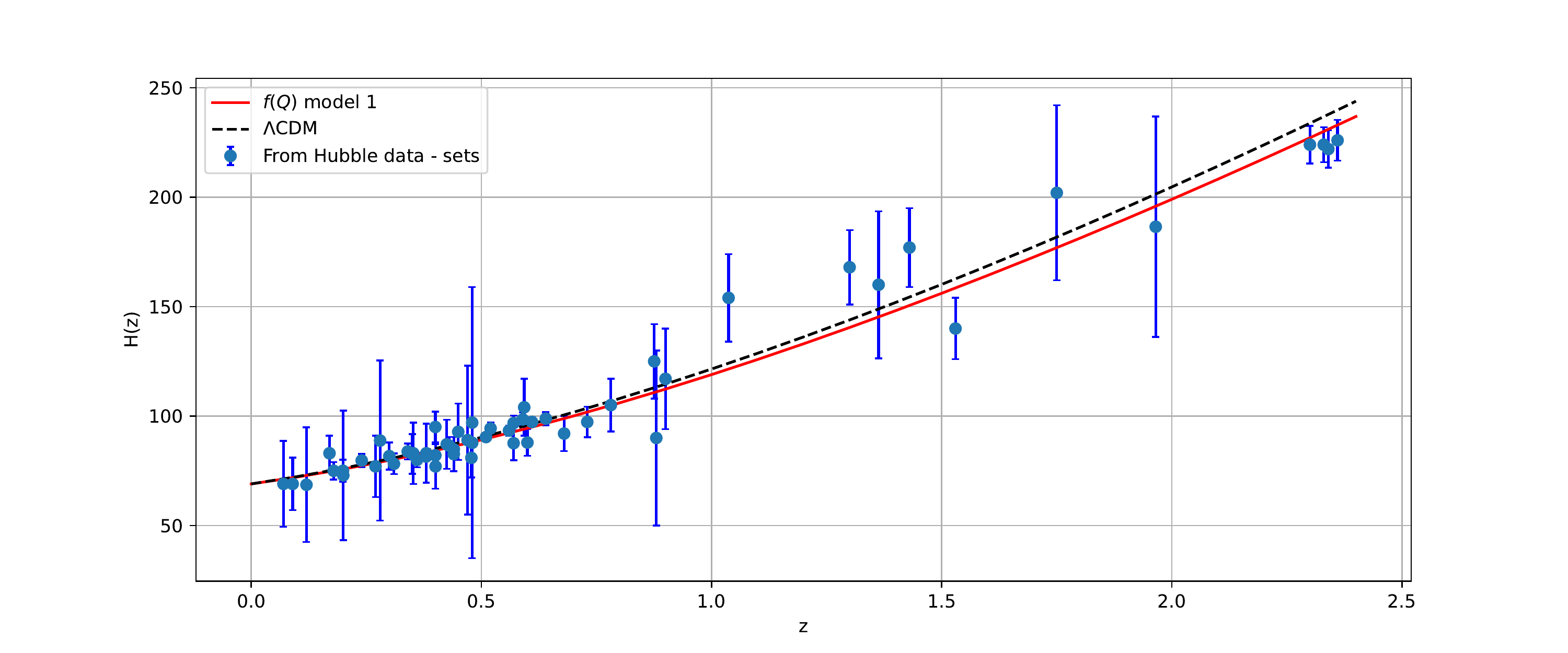}}
\caption{The plot of $H(z)$ vs the red-shift $z$ for our $f(Q)$ model 1, which is shown in red, and $\Lambda$CDM, which is shown in black dashed lines, shows an excellent match to the 57 points of the Hubble datasets.}
\label{ErrorHubble1}
\end{figure}

\begin{figure}[h]
\centerline{\includegraphics[scale=0.60]{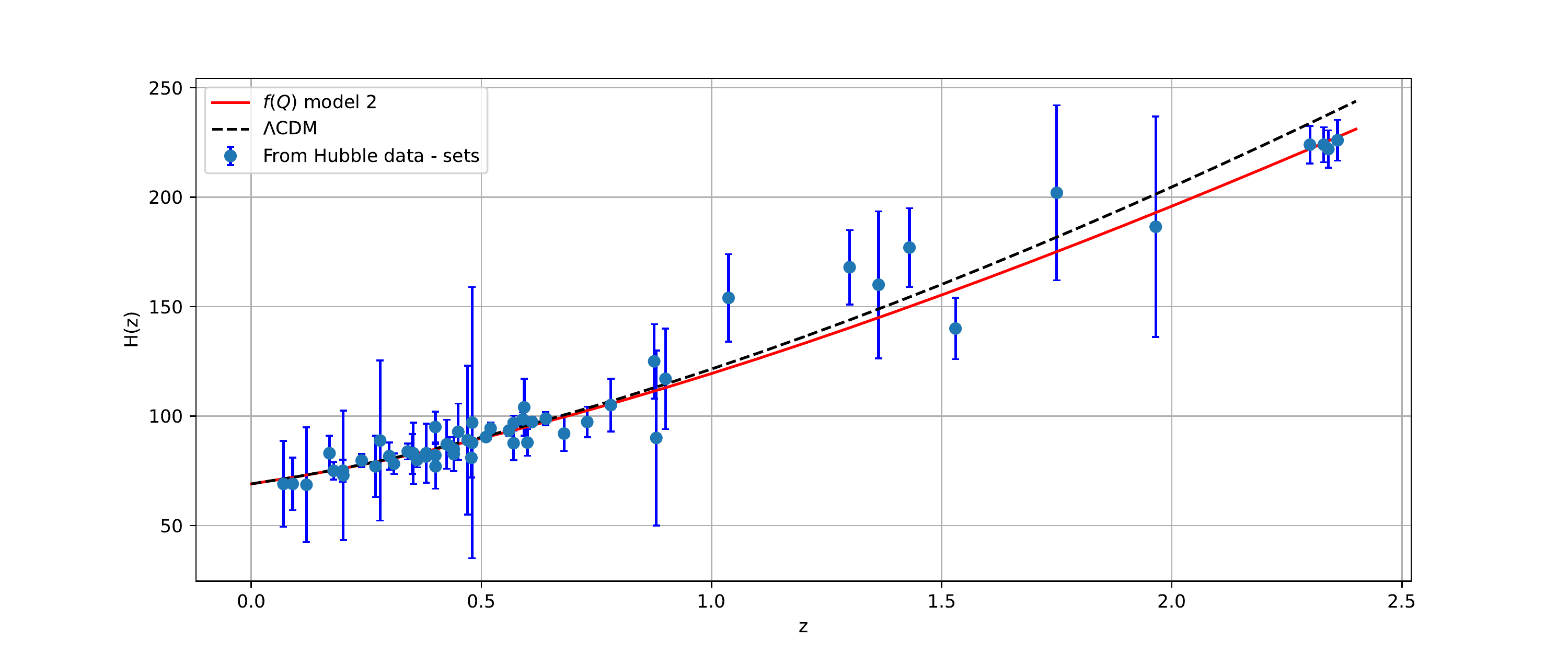}}
\caption{The plot of $H(z)$ vs the red-shift $z$ for our $f(Q)$ model 2, which is shown in red, and $\Lambda$CDM, which is shown in black dashed lines, shows an excellent match to the 57 points of the Hubble datasets.}
\label{ErrorHubble2}
\end{figure}
\end{widetext}

\begin{figure}[!htb]
\begin{minipage}{0.49\textwidth}
     \centering
   \includegraphics[scale=0.6]{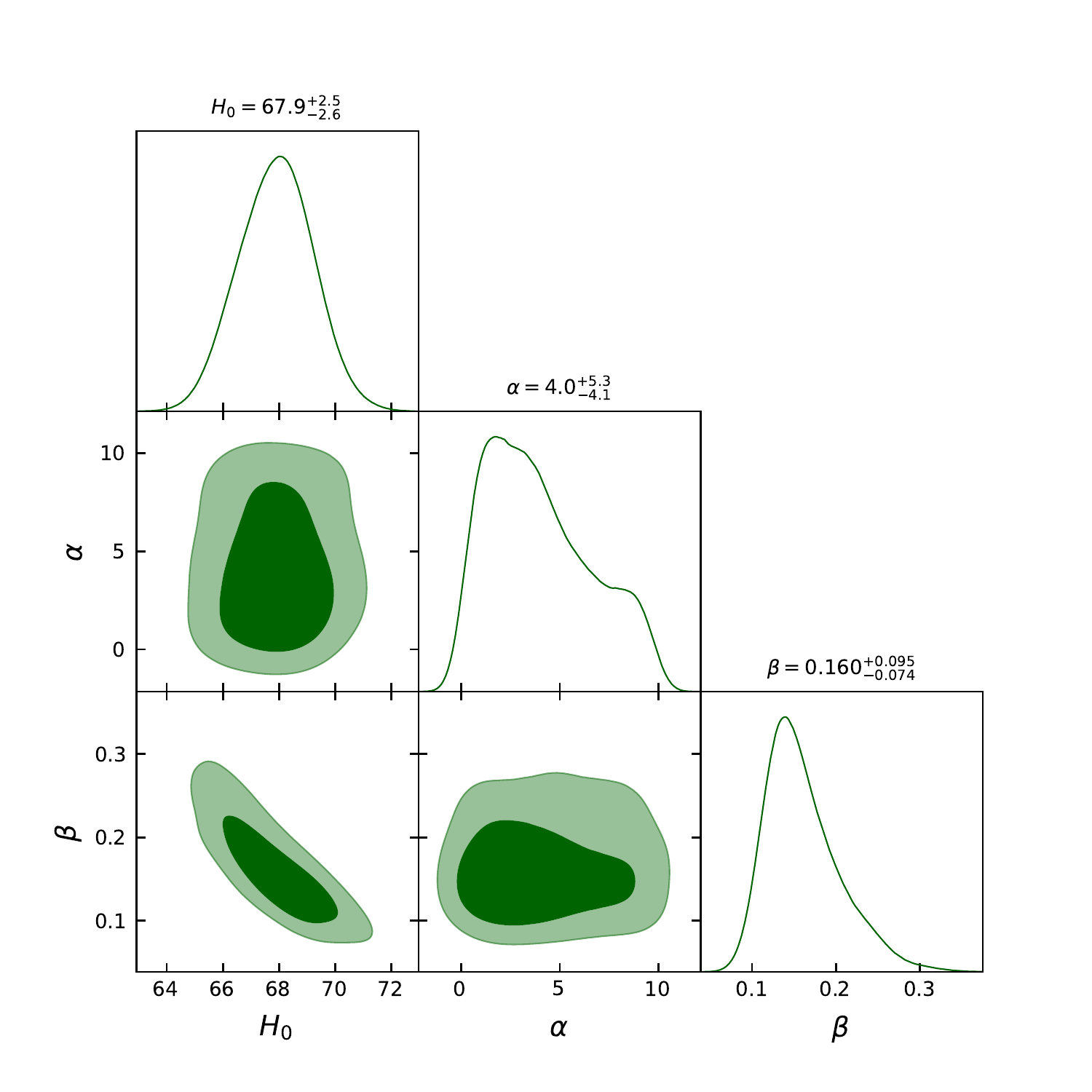}
\caption{Constraints on the model parameters at $1-\protect\sigma $ and $2-%
\protect\sigma $ confidence interval using the SNe Ia datasets (Model 1).}\label{SN1}
   \end{minipage}\hfill 
\begin{minipage}{0.49\textwidth}
     \centering
    \includegraphics[scale=0.6]{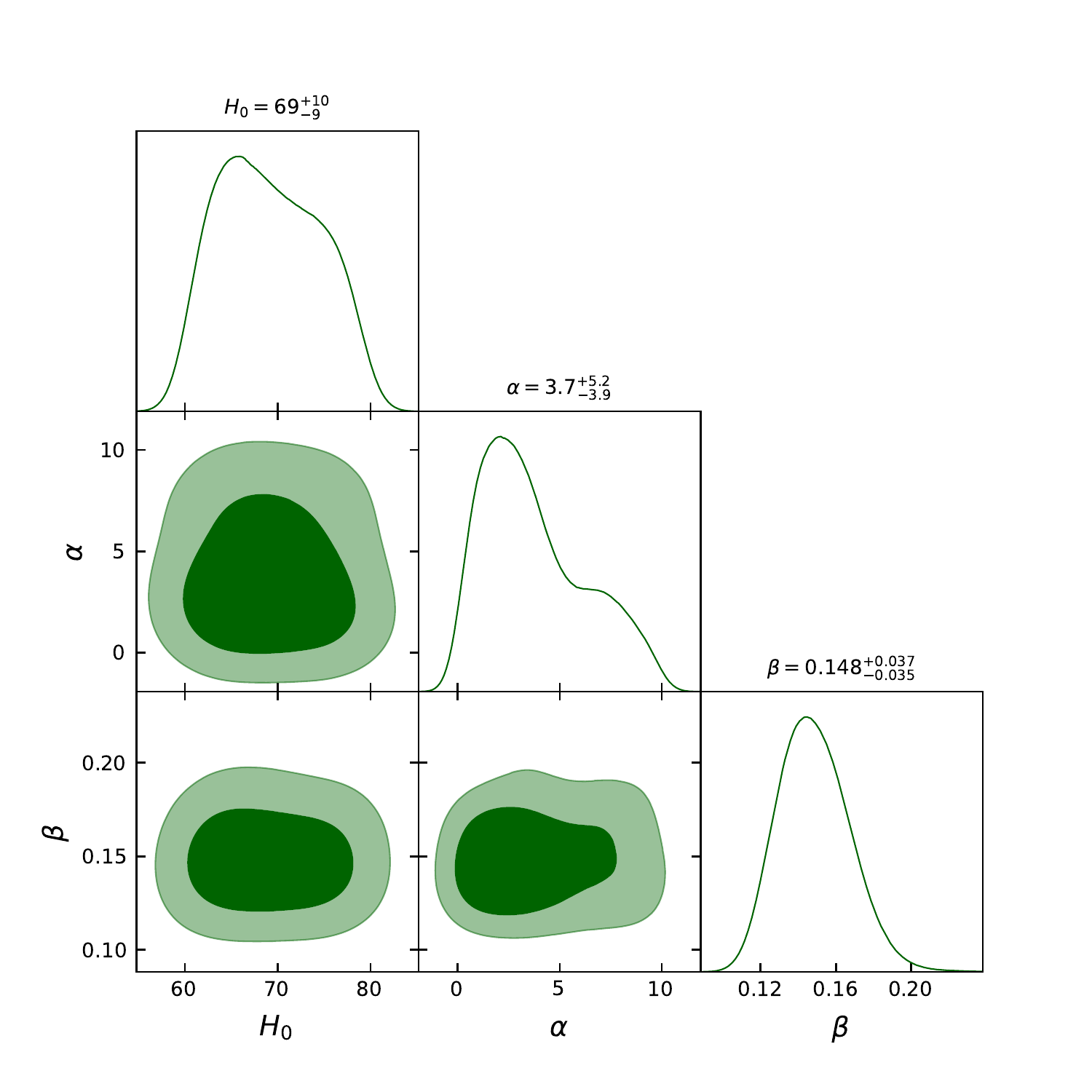}
\caption{Constraints on the model parameters at $1-\protect\sigma $ and $2-%
\protect\sigma $ confidence interval using the BAO datasets (Model 1).}\label{BAO1}
   \end{minipage}
\end{figure}

\subsection{SNe Ia datasets}

Type Ia Supernovae (SNe Ia) are a strong distance indicator that can be
employed to investigate the background expansion of the cosmos. To restrict
the aforementioned parameters, we use the latest Pantheon SNe Ia collection,
which consists of 1048 SNe Ia data points collected from several SNe Ia
samples in the red-shift range $z\in \lbrack 0.01,2.3]$ such as SDSS, SNLS,
Pan-STARRS1, low-red-shift survey, and HST surveys \cite{Scolnic/2018}. For
SNe Ia datasets, the $\chi _{SNe}^{2}$ function is given as 
\begin{equation}
\chi _{SNe\_1}^{2}=\sum_{i,j=1}^{1048}\Delta \mu _{i}\left(
C_{SNe}^{-1}\right) _{ij}\Delta \mu _{j},  \label{4b}
\end{equation}%
where $C_{SNe}$ represents the covariance matrix \cite{Scolnic/2018}, and 
\begin{equation}
\quad \Delta \mu _{i}=\mu ^{th}(z_{i},H_{0},\alpha ,\beta )-\mu
_{i}^{obs},
\end{equation}%
is the difference between the observable distance modulus value from
astronomical data and the theoretical values calculated from the model with
the given parameter $\alpha $, $\beta $, $n$ and $H_{0}$. Furthermore, the
distance modulus is derived as, $\mu =m_{B}-M_{B}$, where $m_{B}$ and $M_{B}$
signify the measured apparent magnitude and absolute magnitude at a given
red-shift $z$ (Trying to retrieve the nuisance parameter using the new BEAMS
with Bias Correction technique (BBC) \cite{BMS}). Its theoretical value is
also given by 
\begin{equation}
\mu (z)=5log_{10}\left[ \frac{D_{L}(z)}{1Mpc}\right] +25,  \label{4d}
\end{equation}%
where 
\begin{equation}
D_{L}(z)=c(1+z)\int_{0}^{z}\frac{dz^{\prime }}{H(z^{\prime },H_{0},\alpha
,\beta )}.
\end{equation}

\begin{widetext}
\begin{figure}[h]
\centerline{\includegraphics[scale=0.60]{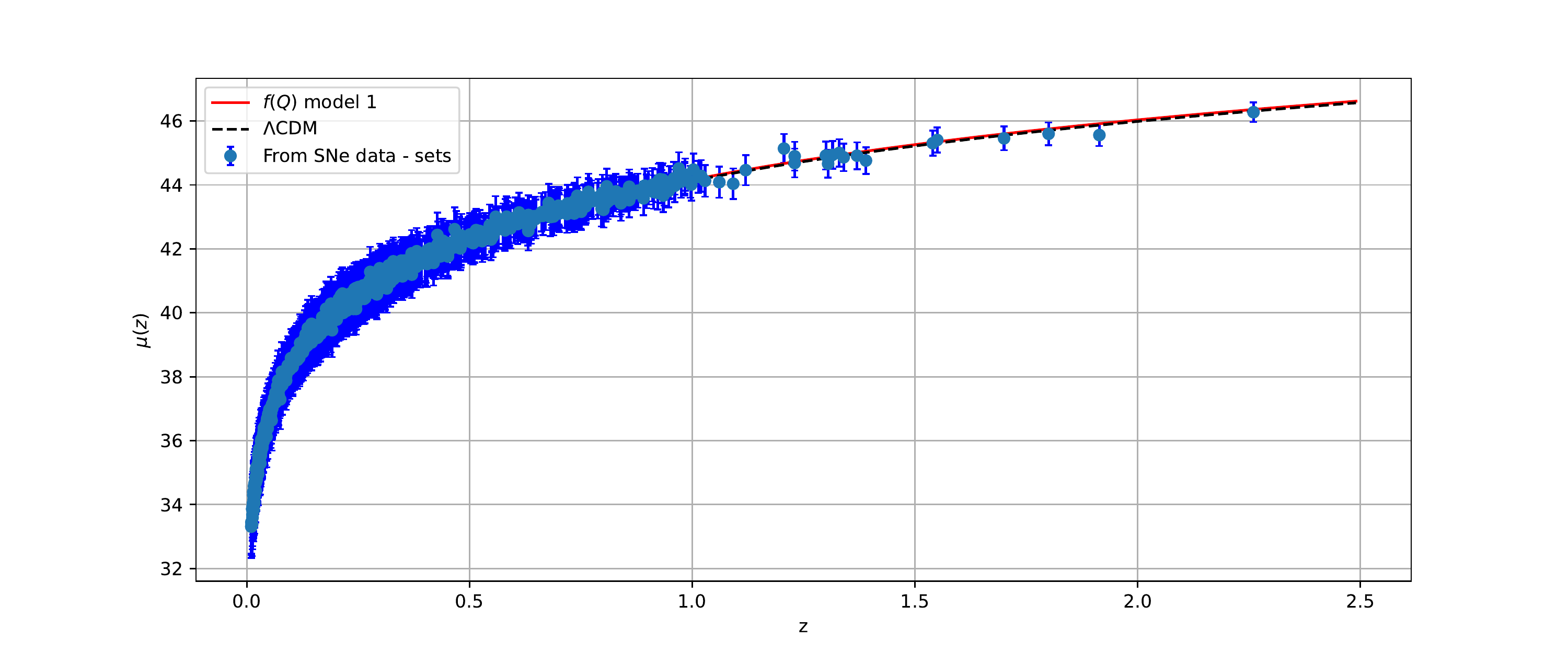}}
\caption{The plot of $\mu(z)$ vs the red-shift $z$ for our $f(Q)$ model 1, which is shown in red, and $\Lambda$CDM, which is shown in black dashed lines, shows an excellent match to the 1048 points of the Pantheon datasets.}
\label{ErrorSNe}
\end{figure}

\end{widetext}

The $1-\sigma $ and $2-\sigma $ contour graphs in Fig. \ref{SN1} show the
best-fit values for the parameters of model 1 obtained from the SNe Ia
datasets. The best-fit values of the model parameters found are: $%
H_{0}=67.9_{-2.6}^{+2.5}$, $\alpha =4.0_{-4.1}^{+5.3}$, and $\beta
=0.160_{-0.074}^{+0.095}$. Furthermore, Fig. \ref{ErrorSNe} shows the error
bar plots for the model and the $\Lambda $CDM, with the cosmological
constant density parameter $\Omega _{\Lambda }^{0}=0.7$, the matter density
parameter $\Omega _{m}^{0}=0.3$ and $H_{0}=69$ km/s/Mpc. The graphic also
displays the SNe Ia findings, 1048 data points, with their errors, allowing
for a direct comparison of the two models.

\subsection{BAO datasets}

The last restrictions in this study are obtained by BAO observation. BAO
studies oscillations in the early cosmos generated by cosmic perturbations
in a fluid composed of photons, baryons, and dark matter and connected via
Thompson scattering. BAO observations include the Sloan Digital Sky Survey
(SDSS), the Six Degree Field Galaxy Survey (6dFGS), and the Baryon
Oscillation Spectroscopy Survey (BOSS) \cite{BAO1,BAO2}. The equations
employed in BAO analysis are 
\begin{equation}
d_{A}(z)=c\int_{0}^{z}\frac{dz^{\prime }}{H(z^{\prime })},  \label{4f}
\end{equation}%
\begin{equation}
D_{V}(z)=\left[\frac{d_{A}^2(z) c z}{H(z)} \right]^\frac{1}{3},  \label{4g}
\end{equation}%
and 
\begin{equation}
\chi _{BAO\_1}^{2}=X^{T}C_{BAO}^{-1}X,  \label{4h}
\end{equation}

\begin{equation*}
X=\left( 
\begin{array}{c}
\frac{d_{A}(z_{\star })}{D_{V}(0.106)}-30.95 \\ 
\frac{d_{A}(z_{\star })}{D_{V}(0.2)}-17.55 \\ 
\frac{d_{A}(z_{\star })}{D_{V}(0.35)}-10.11 \\ 
\frac{d_{A}(z_{\star })}{D_{V}(0.44)}-8.44 \\ 
\frac{d_{A}(z_{\star })}{D_{V}(0.6)}-6.69 \\ 
\frac{d_{A}(z_{\star })}{D_{V}(0.73)}-5.45%
\end{array}%
\right) \,,
\end{equation*}

\begin{widetext}

\begin{table}[H]

\begin{center}

\begin{tabular}{|c|c|c|c|c|c|c|}
\hline
$z_{BAO}$ & $0.106$ & $0.2$ & $0.35$ & $0.44$ & $0.6$ & $0.73$ \\ \hline
$\frac{d_{A}(z_{\ast })}{D_{V}(z_{BAO})}$ & $30.95\pm 1.46$ & $17.55\pm 0.60$
& $10.11\pm 0.37$ & $8.44\pm 0.67$ & $6.69\pm 0.33$ & $5.45\pm 0.31$ \\ 
\hline
\end{tabular}
\caption{Values of $d_{A}(z_{\ast })/D_{V}(z_{BAO})$ for various values of $z_{BAO}$.}
\end{center}
\label{Tab1}
\end{table}

\end{widetext} where $d_{A}(z)$ represents the angular diameter distance, $%
D_{V}(z)$ represents the dilation scale, and $C_{BAO}$ represents the
covariance matrix defined as \cite{BAO6},

\begin{widetext}

\begin{equation*}
C_{BAO}^{-1}=\left( 
\begin{array}{cccccc}
0.48435 & -0.101383 & -0.164945 & -0.0305703 & -0.097874 & -0.106738 \\ 
-0.101383 & 3.2882 & -2.45497 & -0.0787898 & -0.252254 & -0.2751 \\ 
-0.164945 & -2.454987 & 9.55916 & -0.128187 & -0.410404 & -0.447574 \\ 
-0.0305703 & -0.0787898 & -0.128187 & 2.78728 & -2.75632 & 1.16437 \\ 
-0.097874 & -0.252254 & -0.410404 & -2.75632 & 14.9245 & -7.32441 \\ 
-0.106738 & -0.2751 & -0.447574 & 1.16437 & -7.32441 & 14.5022%
\end{array}%
\right) \,.
\end{equation*}

\end{widetext} 

The $1-\sigma $ and $2-\sigma $ contour graphs in Fig. \ref{BAO1} shows the
best-fit values for the parameters of model 2 obtained from the BAO
datasets. The model parameter constraints are derived by minimizing the
associated $\chi ^{2}$ using MCMC and the emcee library. The best-fit values
of the model parameters found are: $H_{0}=69_{-9}^{+10}$, $\alpha
=3.7_{-3.9}^{+5.2}$, and $\beta =0.148_{-0.035}^{+0.037}$.

\section{Behavior of cosmological parameters}\label{section 6}

The behavior of cosmological parameters is determined by the underlying
cosmological model. Several parameters are used to characterize the current
state and evolution of the Universe in the classic $\Lambda $CDM model,
which defines the Universe as spatially flat, homogeneous and isotropic, and
composed of baryonic matter, dark matter, and dark energy such as the
deceleration parameter, energy density, and EoS parameter. In this work, we
study two cosmological models in $f\left( Q\right) $ gravity. Now, we
discuss the behavior of some of the previously mentioned cosmological
parameters.

\begin{figure}[!htb]
\begin{minipage}{0.49\textwidth}
     \centering
   \includegraphics[scale=0.5]{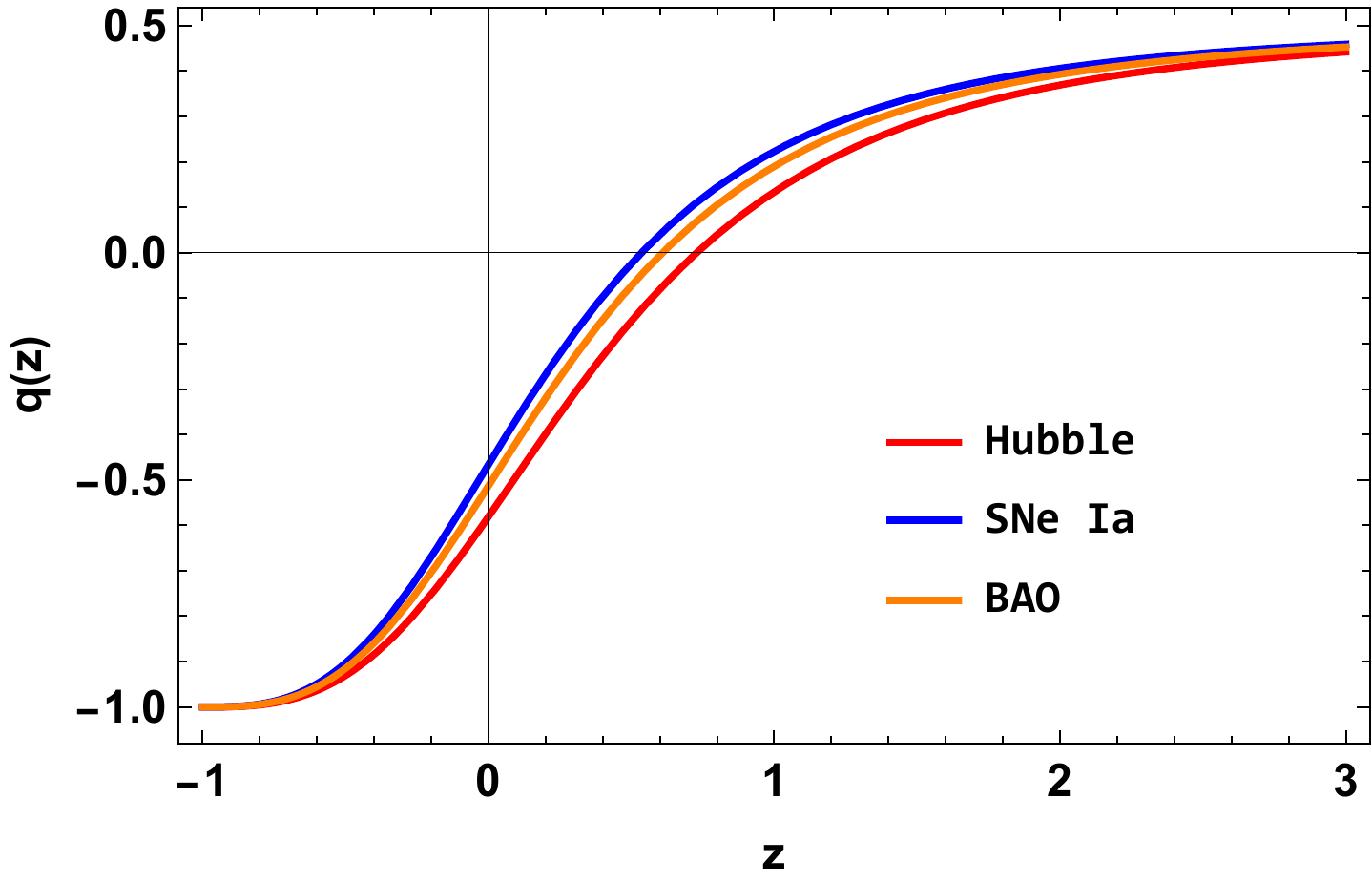}
\caption{Evolution of the deceleration parameter vs red-shift $z$ (Model 1).}\label{q1}
   \end{minipage}\hfill 
\begin{minipage}{0.49\textwidth}
     \centering
    \includegraphics[scale=0.5]{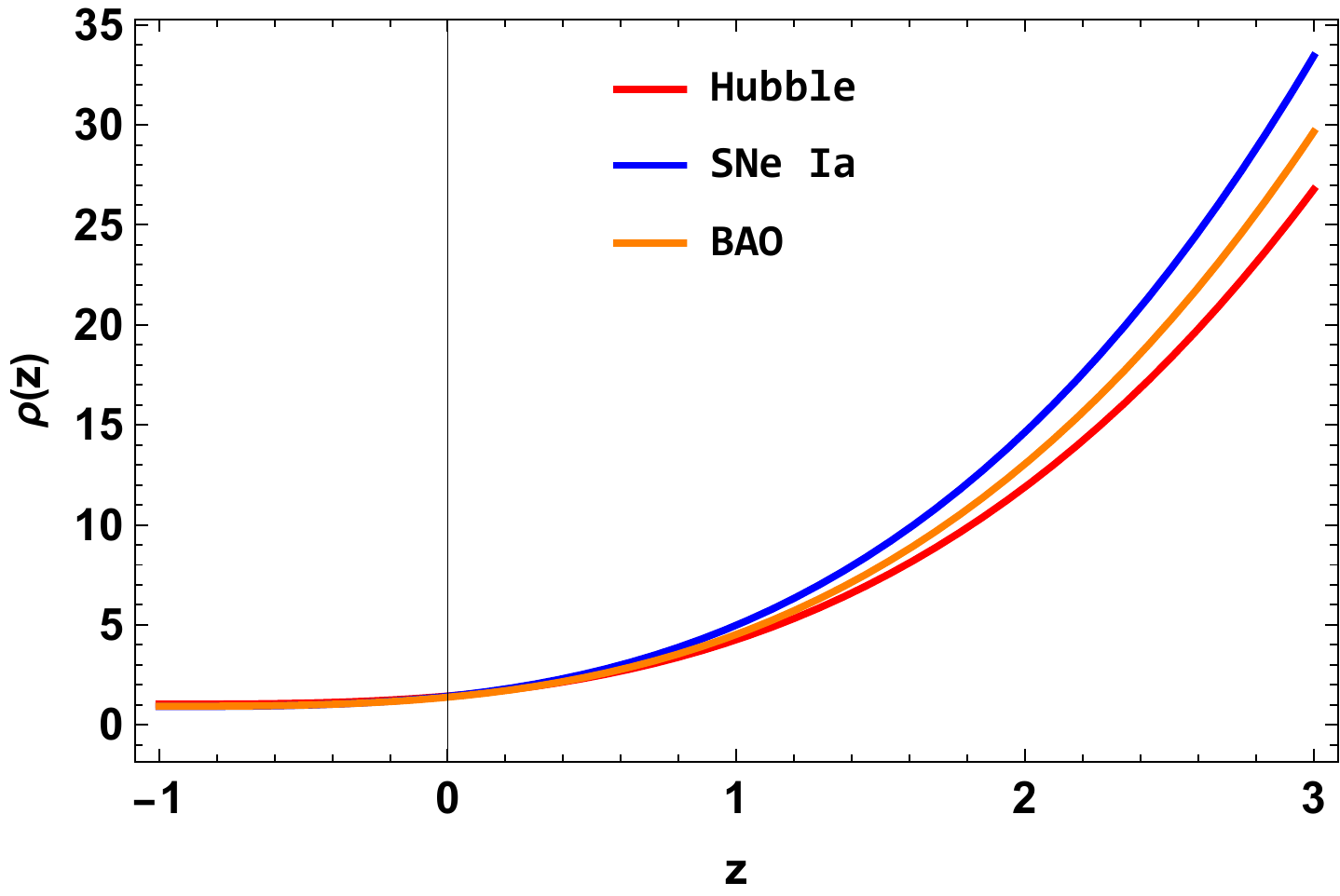}
\caption{Evolution of the energy density vs red-shift $z$ (Model 1).}\label{rho1}
   \end{minipage}
\end{figure}

\begin{figure}[!htb]
\begin{minipage}{0.49\textwidth}
     \centering
   \includegraphics[scale=0.47]{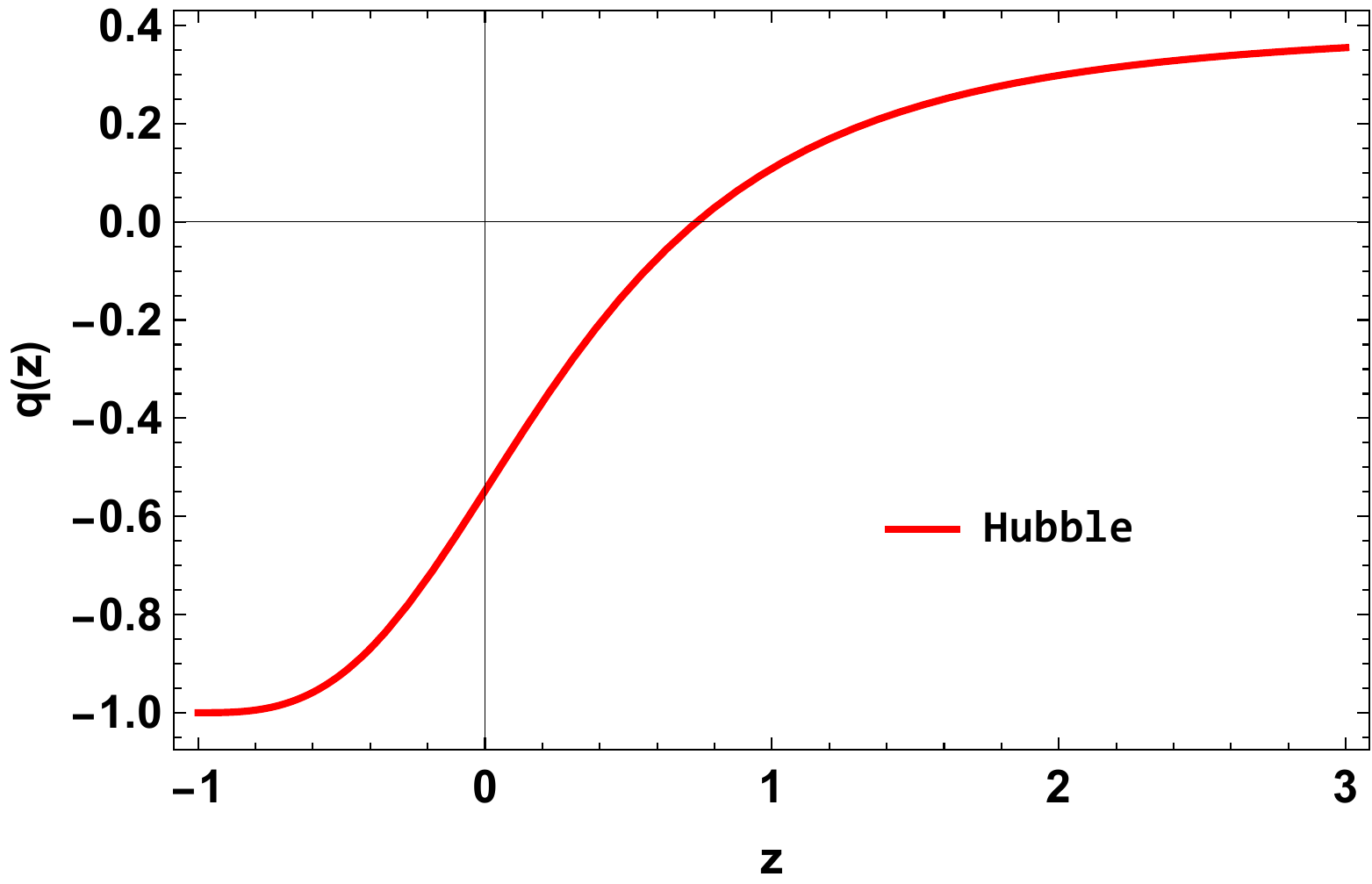}
\caption{Evolution of the deceleration parameter vs red-shift $z$ (Model 2).}\label{q2}
   \end{minipage}\hfill 
\begin{minipage}{0.49\textwidth}
     \centering
    \includegraphics[scale=0.47]{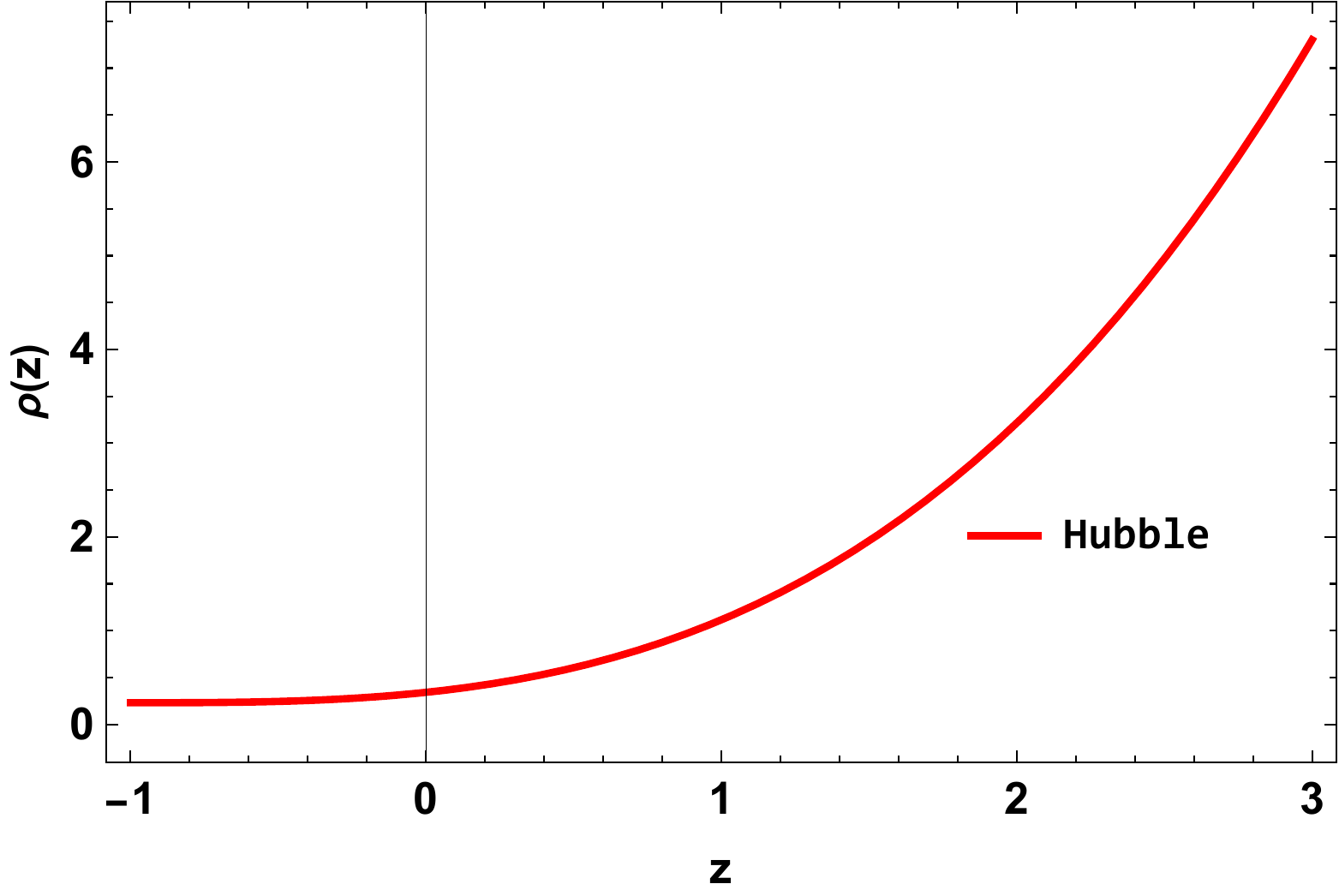}
\caption{Evolution of the energy density vs red-shift $z$ (Model 2).}\label{rho2}
   \end{minipage}
\end{figure}

The deceleration parameter is a measure that estimates the rate of cosmic
expansion. It is obtained as shown in (\ref{q}). The deceleration
parameter's value can be positive, zero, or negative, and it is determined by
the density of matter and the cosmological constant in the Universe. A positive
value of $q$ suggests that the expansion of the Universe is decelerating,
whereas a negative value of $q$ indicates that the expansion is
accelerating. From Figs. (\ref{q1}) and (\ref{q2}), we can observe that the
deceleration parameter of our models is positive ($q>0$) in the early
Universe and negative ($q<0$) in the late cosmos. As a result, it shows that
the cosmos is transitioning from deceleration to acceleration after a
transition red-shift $z_{t}$. Also, the Universe achieves the exponentially
accelerating de Sitter stage with $q=-1$ in the enormous time limit, a
finding that is independent of the model parameters, $q$ decreases as cosmic
time increases or vice versa in terms of red-shift. This evolution is
compatible with the recent Universe's behavior, which passed through three
phases: decelerating dominated, accelerating expansion, and late-time
acceleration. Finally, we observe that the current values of the
deceleration parameter $q_{0}\left( z=0\right) $ and $z_{t}\left( q=0\right) 
$ agree with the Hubble, SNe Ia, and BAO datasets.

\begin{figure}[!htb]
\begin{minipage}{0.49\textwidth}
     \centering
   \includegraphics[scale=0.5]{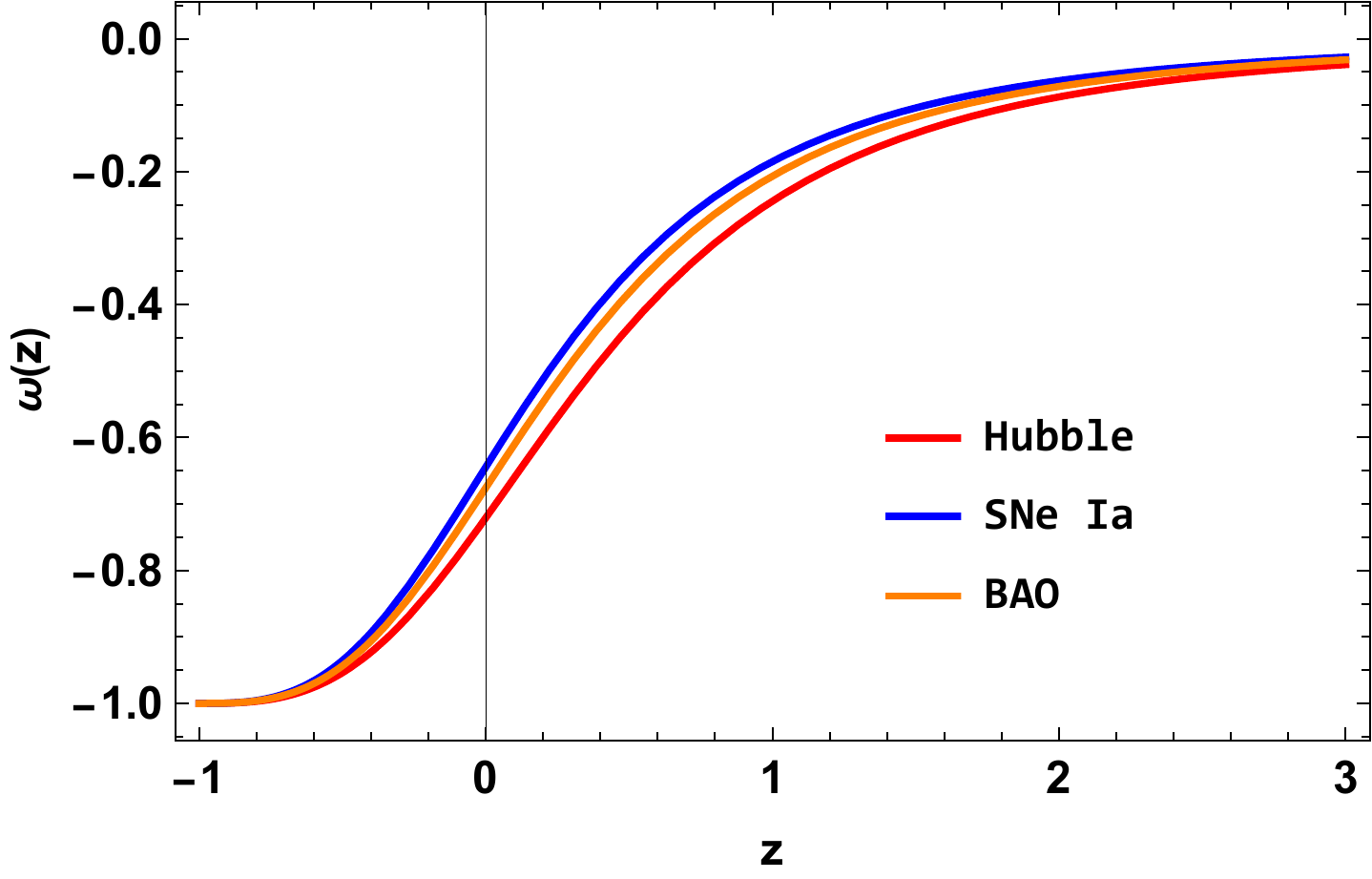}
\caption{Evolution of the EoS parameter vs red-shift $z$ (Model 1).}\label{EoS1}
   \end{minipage}\hfill 
\begin{minipage}{0.49\textwidth}
     \centering
    \includegraphics[scale=0.6]{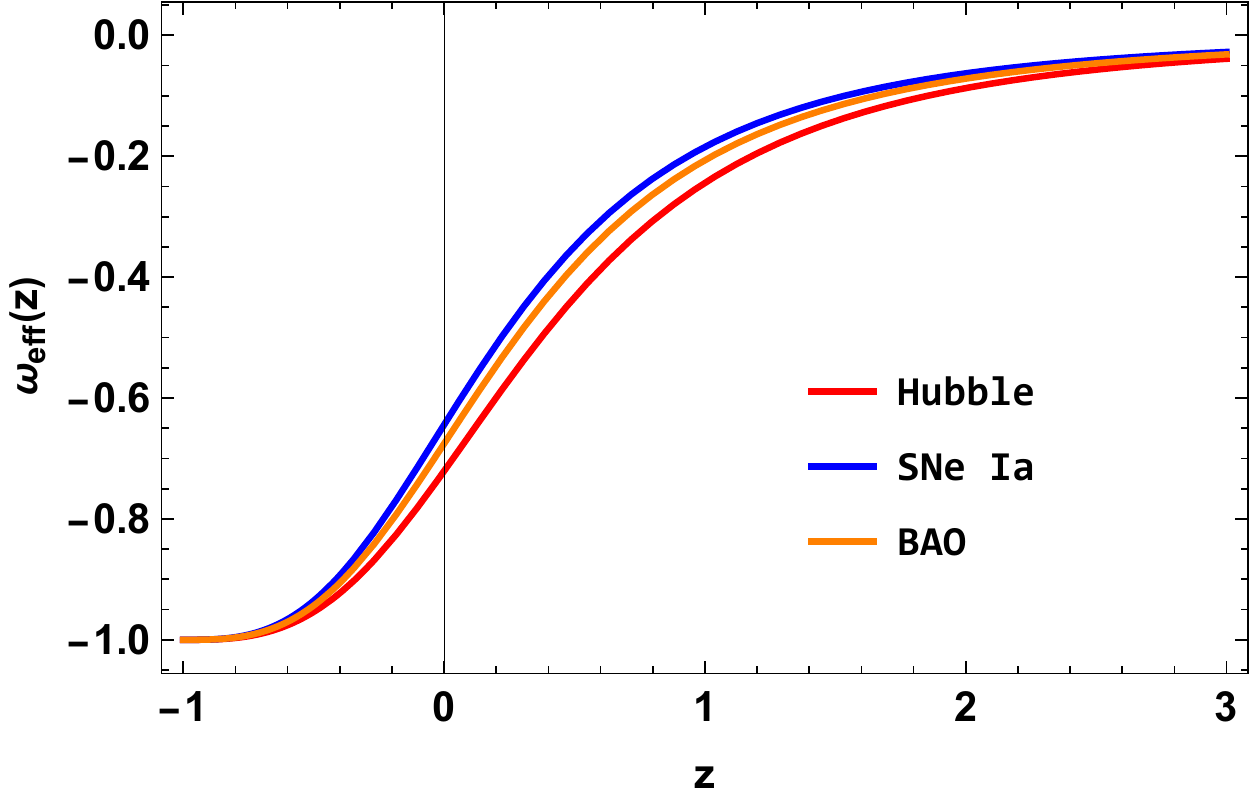}
\caption{Evolution of the effective EoS parameter vs red-shift $z$ (Model 1).}\label{eff1}
   \end{minipage}
\end{figure}

\begin{figure}[!htb]
\begin{minipage}{0.49\textwidth}
     \centering
   \includegraphics[scale=0.47]{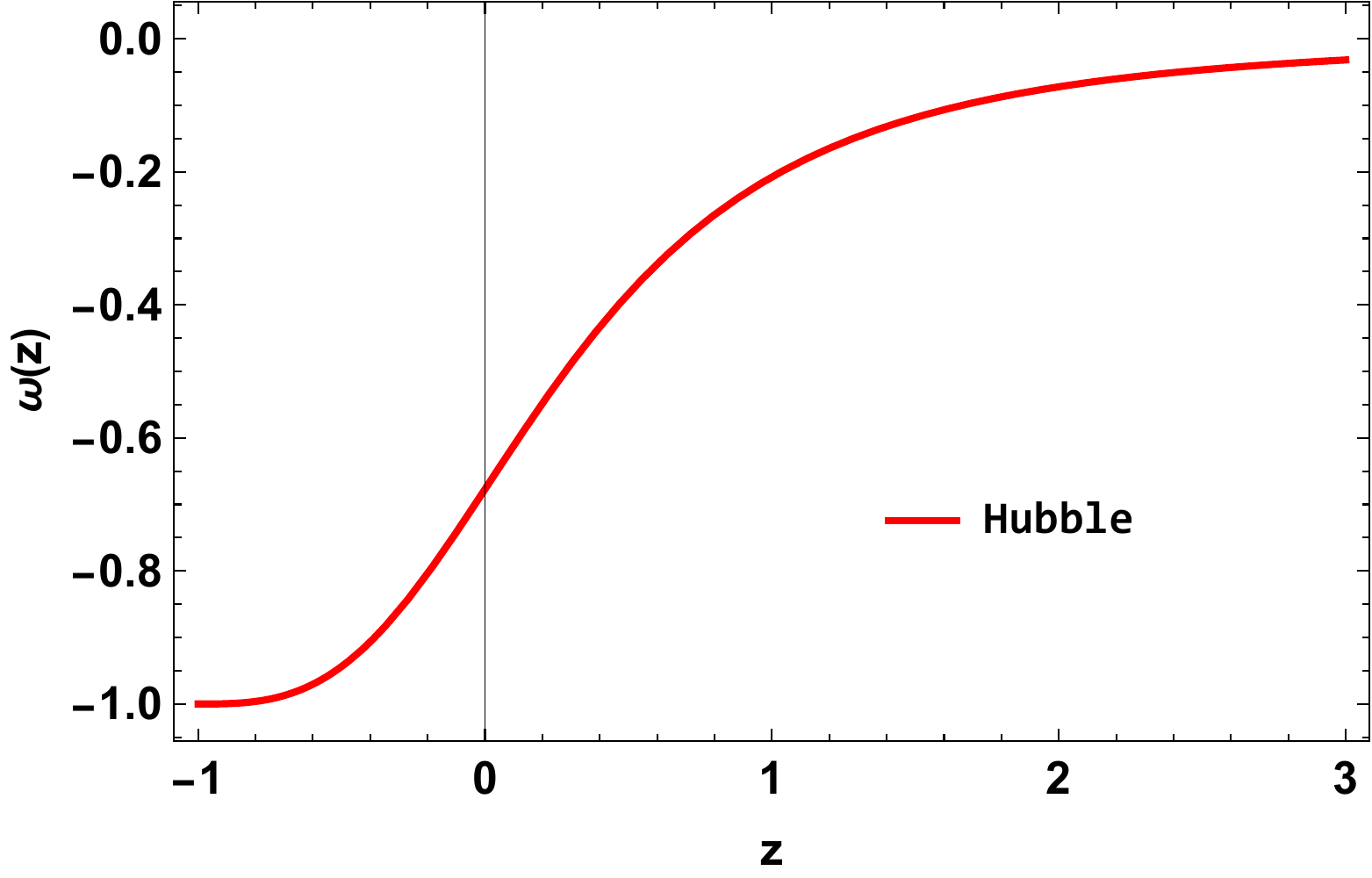}
\caption{Evolution of the EoS parameter vs red-shift $z$ (Model 2).}\label{EoS2}
   \end{minipage}\hfill 
\begin{minipage}{0.49\textwidth}
     \centering
    \includegraphics[scale=0.47]{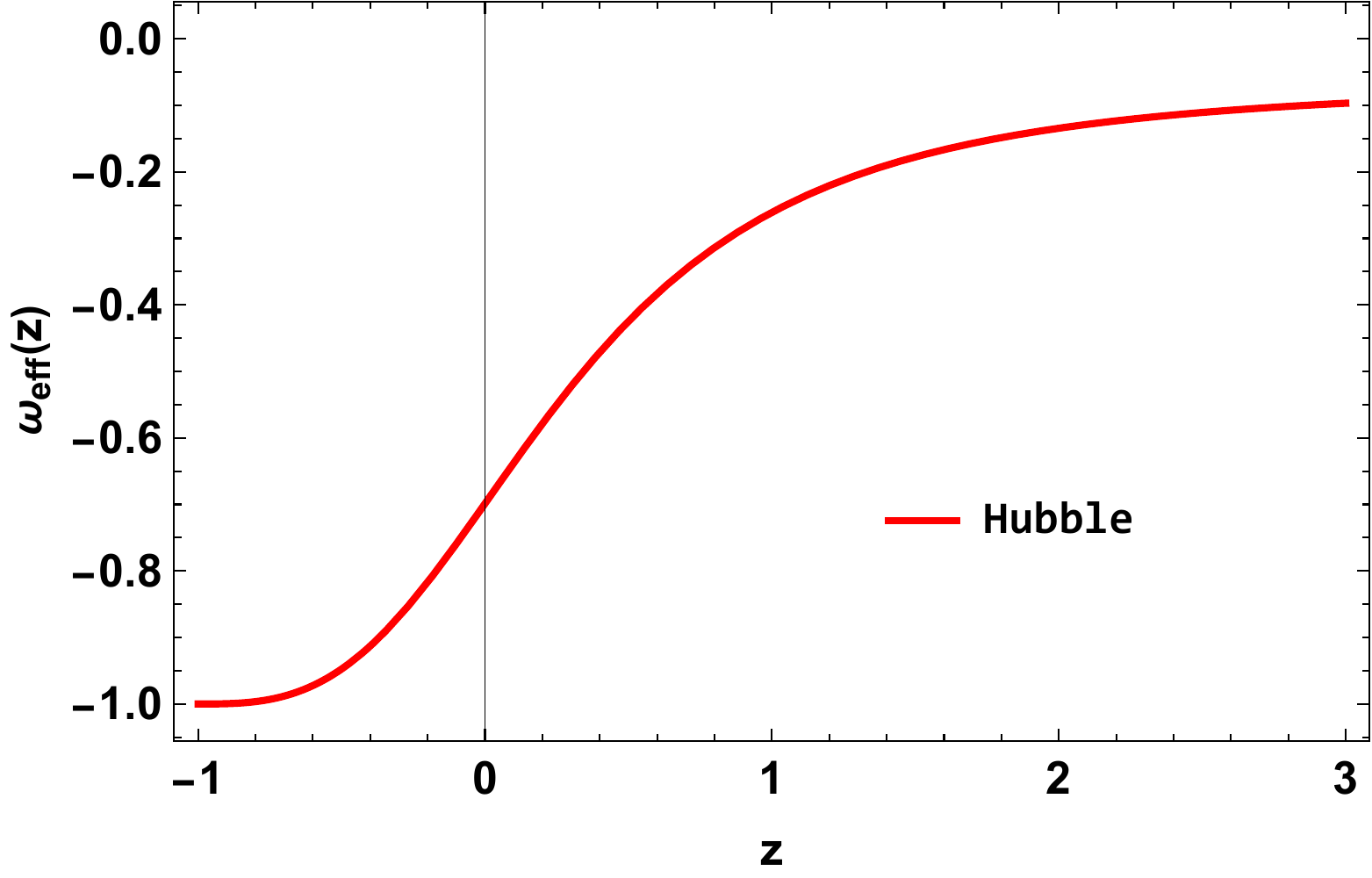}
\caption{Evolution of the effective EoS parameter vs red-shift $z$ (Model 2).}\label{eff2}
   \end{minipage}
\end{figure}

Figs. \ref{rho1} and \ref{rho2} show that the energy density of the cosmos
remains positive all through the Universe's history and decreases as cosmic
time $t$ increases in both models. It starts as positive and decreases to a
small value as $t\rightarrow \infty $ (or $z\rightarrow -1$). Another
attempt to understand the presence of dark energy is to determine the
equation of state (EoS) value and its evolution. Figs. \ref{EoS1} and \ref%
{EoS2} show the behavior of the EoS parameter for two models. So, the EoS
parameter is a dimensionless quantity that represents the pressure-to-energy
density ratio in a cosmic fluid i.e. $\omega =\frac{p}{\rho }$. It is
frequently used in cosmology to explain the behavior of dark energy and dark
matter, which are considered to make up the majority of the Universe. The
EoS parameter can also have values varying from $-1$ to $1$. A fluid with a
value of $\omega =-1$ behaves like a cosmological constant, such as dark
energy, whereas a fluid with a value of $\omega =0$ behaves like
non-relativistic matter, such as dark matter. The effective EoS parameter in
Figs. \ref{eff1} and \ref{eff2} appears to be similar, takes negative values
for all values of $z$. As a result, both EoS are in the quintessence region (%
$-1<\omega <-\frac{1}{3}$), approaching the cosmological constant at high
red-shifts. Finally, we observe that the current values of the EoS parameter $%
\omega _{0}\left( z=0\right) $ agree with the Hubble, SNe Ia, and BAO
datasets.

\section{Concluding remarks}\label{section 7}

The present era of accelerating Universe has gotten increasingly intriguing
over time. To develop a proper description of the accelerating Universe, a
variety of dynamical DE models and modified gravity models have been used in
different ways. We have investigated the accelerated expansion of the
Universe in this paper by using the parametric form of the equation of state
parameter in the context of $f(Q)$ gravity, where the gravitational
interaction is represented by the non-metricity scalar $Q$. We have studied
two functional forms of $f(Q)$, specifically $f(Q)=-Q+\frac{\alpha }{Q}$ and 
$f(Q)=-\alpha Q^{n}$, where $\alpha $ and $n$ are model parameters, and the
parametrization form of the equation of state parameter as $\omega \left(
z\right) =-\frac{1}{1+3\beta \left( 1+z\right) ^{3}}$, where $\beta $ is a
constant parameter.

Using the aforementioned parametric form, we can derive the Hubble parameter
as shown in (\ref{H1}) and (\ref{H2}). In addition, we have utilized Hubble
datasets with 57 data points, SNe Ia datasets with 1048 data points, and BAO
datasets with six data points to find the best-fit values for the model
parameter. The values for the first model are as: $H_{0}=69.3_{-2.2}^{+2.4}$%
, $\alpha =3.9_{-4.0}^{+5.2}$, and $\beta =0.129_{-0.024}^{+0.027}$ for
Hubble datasets, $H_{0}=67.9_{-2.6}^{+2.5}$, $\alpha =4.0_{-4.1}^{+5.3}$,
and $\beta =0.160_{-0.074}^{+0.095}$ for SNe Ia datasets, and $%
H_{0}=69_{-9}^{+10}$, $\alpha =3.7_{-3.9}^{+5.2}$, and $\beta
=0.148_{-0.035}^{+0.037}$ for BAO datasets. In the case of the second model: 
$H_{0}=65.5_{-4.6}^{+4.5}$, $n=1.16_{-0.15}^{+0.14}$, and $\beta
=0.33_{-0.24}^{+0.39}$ for Hubble datasets. We have examined the behavior of
several cosmological parameters for both models according to these best-fit
model parameter values. The deceleration parameter behavior of both models
has predicted the phase transition from deceleration ($q>0$) to acceleration
($q<0$). This shows that the current expansion of the Universe is
accelerating. Furthermore, the effective EoS parameter in both models
presently behaves in the same method as the quintessence behavior, i.e.$%
-1<\omega <-\frac{1}{3}$, with the current values in the negative range and
close to the observed value. It shows that the current Universe is
accelerating. Finally, we have concluded that the proposed parameterization
form of the EoS parameter in the context of $f(Q)$ gravity theory plays a
significant role in demonstrating the late-time accelerated expansion of the
Universe.

\textbf{Data availability} There are no new data associated with this
article.

\textbf{Declaration of competing interest} The authors declare that they
have no known competing financial interests or personal relationships that
could have appeared to influence the work reported in this paper.\newline


\end{document}